\documentclass[structabstract]{aa}  
\usepackage{graphicx}
\usepackage{txfonts}
\usepackage{longtable}
\begin{document}
   \title{The galactic unclassified B[e] star HD\,50138\thanks{Based on observations (i) with the 1.52-m and 2.2-m telescopes at the European Southern Observatory (La Silla, Chile), under agreement with the Observat\'orio Nacional-MCT (Brazil), and (ii) with the Telescope Bernard Lyot, Observatory of Pic du Midi (France).}}
   \subtitle{I. A possible new shell phase}

   \author{M. Borges Fernandes
          \inst{1},
          M. Kraus
          \inst{2},
          O. Chesneau
          \inst{1},
          A. Domiciano de Souza
          \inst{3},
          F. X. de Ara\'ujo\thanks{It is with great sadness that we have to report that, during the final stages of this paper, we had a deep loss when Francisco X. de Ara\'ujo passed away.}
          \inst{4},
	  P. Stee
          \inst{1},
          \and
          A. Meilland
          \inst{5}
          }

   \institute{UMR 6525 H. Fizeau, Univ. Nice Sophia Antipolis, CNRS, Observatoire de
la C\^{o}te d'Azur, Av. Copernic, 06130 Grasse, France\\
              \email{marcelo.borges@obs-azur.fr, olivier.chesneau@obs-azur.fr, philippe.stee@obs-azur.fr}
         \and
             Astronomick\'y \'ustav, Akademie v\v{e}d \v{C}esk\'e republiky, Fri\v{c}ova 298, 251\,65 Ond\v{r}ejov, Czech Republic\\
             \email{kraus@sunstel.asu.cas.cz}
         \and
             UMR 6525 H. Fizeau, Univ. Nice Sophia Antipolis, CNRS, Observatoire de
la C\^{o}te d'Azur, Parc Valrose, 06108 Nice, France\\
             \email{armando.domiciano@unice.fr}
         \and
             Observat\'orio Nacional, Rua General Jos\'e Cristino 77, 20921-400 S\~ao Cristov\~ao, Rio de Janeiro, Brazil\\
             \email{araujo@on.br}
         \and
             Max-Planck-Institut f\"ur Radioastronomie, Auf dem H\"ugel 69, 53121 Bonn, Germany\\
             \email{meilland@mpifr-bonn.mpg.de} 
             }

   \date{Received; accepted}

\authorrunning{Borges Fernandes et al.}
\titlerunning{The galactic unclassified B[e] star HD\,50138}

 
  \abstract
   {The observed spectral variation of HD\,50138 has led different authors to classify it in a very wide range of spectral types and luminosity classes (from B5 to A0 and III to Ia) and at different evolutionary stages as either HAeBe star or classical Be.  }
   {Based on new high-resolution optical spectroscopic data from 1999 and 2007 associated to a photometric analysis, the aim of this work is to provide a deep spectroscopic description and a new set of parameters for this unclassified southern B[e] star and its interstellar extinction.     }
   {From our high-resolution optical spectroscopic data separated by 8 years, we perform a detailed spectral description, presenting the variations seen and discussing their possible origin. We derive the interstellar extinction to HD\,50138 by taking the influences of the circumstellar matter in the form of dust and an ionized disk into account. Based on photometric data from the literature and the new Hipparcos distance, we obtain a revised set of parameters for HD\,50138.}
   {Because of the spectral changes, we tentatively suggest that a new shell phase could have taken place prior to our observations in 2007. We find a color excess value of E(B-V) = 0.08\,mag, and from the photometric analysis, we suggest that HD\,50138 is a B6-7 III-V star. A discussion of the different evolutionary scenarios is also provided.    }
   {}


   \maketitle
%

\section{Introduction}

Stars that present the B[e] phenomenon are known to form a heterogeneous group. This group is composed of objects at different evolutionary stages, such as high- and low-mass evolved stars, intermediate-mass pre-main sequence stars, and symbiotic objects (Lamers et al. \cite{Lamers}). However, for more than 50\% of the confirmed galactic B[e] stars, the evolutionary stage is still unknown, so that they are gathered in the group of the unclassified B[e] stars. This problem is mainly caused by poor knowledge of their physical parameters, especially their distances. 

In this paper, we present our study related to the southern B[e] star \object{HD\,50138} (V743 Mon, MWC158, IRAS 06491-0654). The spectrum of this star was discussed for the first time by Merrill et al. (\cite{Merrill1}). Later, the presence of spectral variability was cited by Merrill (\cite{Merrill2}) and Merrill \& Burwell (\cite{Merrill3}). Since then, numerous papers have mainly considered this star as either a pre-main sequence star, more specifically as a Herbig Ae/Be star, or as a classical Be star. Jaschek et al. (\cite{Jaschek2}) have even considered this star as a transition object between a classical Be and a B[e] star. Later Lamers et al. (\cite{Lamers}) and Zorec (\cite{Zorec}) include this star in the list of unclassified B[e] stars. 

The difficulty in obtaining the correct classification of HD\,50138 is mainly caused by the strong spectral variability as reported in the literature. Pogodin (\cite{Pogodin}) cites the presence of different scales of spectral variability, from days to months, especially in the line profiles of H$\alpha$, He{\sc i} ($\lambda$ 5876), and Na{\sc i} lines. Spectral variations have also been identified in the UV, by the analysis of IUE spectra (Hutsem\'ekers \cite{Hutsemekers}). These spectral variabilities have been explained by an outburst that possibly happened in 1978-1979 (Hutsem\'ekers \cite{Hutsemekers}) and by a shell phase in 1990-1991 (Andrillat \& Houziaux \cite{Andrillat}). On the other hand, HD\,50138 had presented small and non-periodic photometric variations that did not seem to be associated to the spectral changes related to this shell phase (Halbedel \cite{Halbedel}).

In addition, studies based on polarimetry and spectro-polarimetry have identified an intrinsic polarization that seems to be linked to nonspherical symmetry of matter around this object, probably a circumstellar disk (Vaidya et al. \cite{Vaidya}; Bjorkman et al. \cite{Bjorkman}; Oudmaijer \& Drew \cite{Oudmaijer}; Harrington \& Kuhn \cite{Harrington}). A disk-scattering effect associated to an outflow has also recently been suggested by Harrington \& Kuhn (\cite{Harrington2}).

On the other hand, Cidale et al. (\cite{Cidale}) propose that this object could be a binary system. Based on spectro-astrometry, Baines et al. (\cite{Baines}) suggest the same possibility, where the companion would be separated by 0.5$\arcsec$ - 3.0$\arcsec$. However, up to now, no direct evidence of binarity has been found.

HD\,50138 was further observed during the Hipparcos mission, and its distance of 290$\pm$71\,pc (Perryman et al. \cite{Perryman}) turned out with 500$\pm$150\,pc to be almost doubled according to the new data reduction procedure performed by van Leeuwen (\cite{vanLeeuwen}). The newly determined distance and the large uncertainties in the
previously published stellar classification attempt request and warrant detailed
investigation and revision of the stellar parameters of HD\,50138.

In this study, we present our new optical spectroscopic observations and perform
a photometric and spectroscopic analysis of HD\,50138, aimed on the one hand at describing the observed spectral variations and, on the other, at better constraining the 
stellar parameters, needed for an improved discussion about the nature of this 
object. 

The paper has the following structure. In Sect.\,\ref{obs} we describe our observations. In Sect.\,\ref{results}, we present our results. In Sect.\,\ref{spectral_descr}, we describe our high-resolution spectra taken on different dates. In Sect.\,\ref{SpType}, we derive from the analysis of our spectroscopic data and public photometric measurements, the insterstellar and circumstellar extinction, hence the stellar parameters and the spectral type of this object. In Sect.\,\ref{discussion}, we discuss the possible scenarios for explaining the nature of this curious star, and finally in Sect.\,\ref{conclusion}, we present our conclusions.


\section{Observations}\label{obs}

We obtained high-resolution optical spectra using the high-resolution Fiber-fed Extended Range Optical Spectrograph 
(FEROS) and the Narval spectro-polarimeter. The first FEROS spectrum was obtained on October 27, 1999, when the spectrograph was attached to the 1.52-m telescope, and the second was obtained on October 4, 2007, when it was attached to the 2.2-m telescope, both at the European Southern Observatory in La Silla (Chile). FEROS is a bench-mounted Echelle spectrograph with fibers, which covers a sky area of 2$\arcsec$ of diameter, with a wavelength coverage from 3600\,\AA \ to 9200\,\AA \ and a spectral resolution of R = 55\,000 (in the region around 6000 
\AA). We adopted its complete automatic online reduction, where the heliocentric correction is done.

The spectrum of 1999 was obtained with an exposure time of 180 seconds and has S/N of approximately 80 in the 5500\,\AA \ region. In 2007, we were able to take two consecutive spectra of the star, both with 180 seconds of exposure time. Since these spectra do not show significant differences, we added them up for a better S/N, which is around 250.

Concerning the Narval data, they were obtained on March 14, 2007. Narval is an Echelle spectro-polarimeter attached to the telescope Bernard Lyot at the observatory of Pic du Midi (France). For this study, we are only using the spectroscopic data that cover a sky area of 2.8$\arcsec$ of diameter, with a wavelength range from 3750\,\AA \ to 10500\,\AA \ and a spectral resolution of R = 80\,000. We obtained 8 exposures of 300 seconds each. The S/N is around 360. We also adopted its complete automatic online reduction; however, due to problems related to the merging of the spectral orders (especially in the region of the Balmer lines), we use this spectrum mainly for a qualitative comparison with the FEROS data, which are of better quality for our purposes.


\section{Results}\label{results}

\subsection{Spectral description of HD\,50138}\label{spectral_descr}

The high-resolution spectra of HD\,50138 present lines from neutral and singly ionized elements. In addition to many emission lines with circumstellar origin, HD\,50138 exhibits absorption lines, which are probably formed in the stellar photosphere. Our spectra were taken 8 years apart and strong spectral variations can be noted. Similar variations were previously reported in the literature and associated to shell phases and outburst events (Doazan \cite{Doazan}; Hutsem\'ekers \cite{Hutsemekers}; Andrillat \& Houziaux \cite{Andrillat}; Bopp \cite{Bopp}; Pogodin \cite{Pogodin}).

Table\,\ref{identification} lists all lines present in our FEROS spectra, along with their radial velocities (obtained considering the center of the lines and the laboratory wavelength of each transition), equivalent widths of emission and absorption components, and the possible identification\footnote{In order to identify the lines, we used the line lists provided by Moore (\cite{Moore}), Thackeray (\cite{Thackeray}), and Landaberry et al. (\cite{Landaberry}). We also looked up two sites on the web: NIST Atomic Spectra Database Lines Form (URL physics.nist.gov/cgi-bin/AtData/lines\_form) and The Atomic
Line List v2.04 (URL http://www.pa.uky.edu/$\sim$peter/atomic).}. For some lines there are more than one possible classification, however the radial velocity is derived assuming the first identification cited. There, ``Uid" means that the line is unidentified. Because of uncertainties in the position of the underlying continuum, we estimate the errors of our measurements to be about 20\% for the faint lines and about 10\% for the strongest lines. Especially for the faint lines, there is also a significant difference between the radial velocities measured in 1999 and in 2007. This difference probably comes not only from circumstellar changes, but also from the low S/N of our 1999 data. The behavior of the lines from the main elements present in our spectra is described here.

\onllongtab{1}{
\begin{longtable}{ccccccccccc}
\caption{Lines identified in our FEROS spectra obtained in 1999 and 2007. The observed central wavelength ($\lambda_{obs}$), radial velocity ($\varv_{\rm rad}$), equivalent width of emission ($W_{em}$) and/or absorption ($W_{abs}$) components and identification of the lines seen in each date are provided.} \\
\hline
\multicolumn{5}{c}{\textrm \protect{1999}} & \multicolumn{5}{c}{\textrm \protect{2007}} & \\
 \hline
 $\lambda$$_{obs}$ & $\varv_{\rm rad}$ & $W_{em}^{blue}$ & $W_{abs}$ & $W_{em}^{red}$ & $\lambda$$_{obs}$ & $\varv_{\rm rad}$ & $W_{em}^{blue}$ & $W_{abs}$ & $W_{em}^{red}$  & Identification \\
 \hline
\endfirsthead
\caption{Continued.} \\
\hline
\multicolumn{5}{c}{\textrm \protect{1999}} & \multicolumn{5}{c}{\textrm \protect{2007}} & \\
 \hline
 $\lambda$$_{obs}$ & $\varv_{\rm rad}$ & $W_{em}^{blue}$ & $W_{abs}$ & $W_{em}^{red}$ & $\lambda$$_{obs}$ & $\varv_{\rm rad}$ & $W_{em}^{blue}$ & $W_{abs}$ & $W_{em}^{red}$  & Identification \\
\hline
\endhead
\hline
\endfoot
\hline
\endlastfoot
  &    &   &   &   & 3704.3 & 32.4  &      & 0.79 &      & H16 3703.9 \\
  &    &   &   &   & 3712.4 & 32.3  &      & 1.07 &      & H15 3712.0 \\
  &    &   &   &   & 3722.5 & 48.4  &      & 1.48 &      & H14 3721.9 \\
  &    &   &   &   & 3735.0 & 48.2  &      & 2.46 &      & H13 3734.4 \\
  &    &   &   &   & 3750.9 & 56.0  &      & 3.37 &      & H12 3750.2 \\
  &    &   &   &   & 3771.4 & 63.7  &      & 4.87 &      & H11 3770.6 \\
  &    &   &   &   & 3798.6 & 55.3  &      & 5.45 &      & H10 3797.9 \\
  &    &   &   &   & 3820.4 & 47.1  &      & 0.22 &      & He\,{\sc i} (m22) 3819.8 \\
  &    &   &   &   &         &        &      &      &      & He\,{\sc i} (m22) 3819.6 \\
  &    &   &   &   & 3835.9 & 39.1  &      & 5.81 &      & H9 3835.4 \\
  &    &   &   &   & 3850.0 &   &      & 0.04 &      & Uid \\
  &    &   &   &   & 3854.0 & 23.4  &      & 0.05 &      & Si\,{\sc ii} (m1) 3853.7  \\
  &    &   &   &   & 3856.5 & 38.9  &      & 0.19 &      & Si\,{\sc ii} (m1) 3856.0  \\
  &    &   &   &   & 3863.1 & 38.8  &      & 0.15 &      & Si\,{\sc ii} (m1) 3862.6  \\
  &    &   &   &   & 3868.0 & 38.8  &      & 0.04 &      & He\,{\sc i} (m20) 3867.6 \\
  &    &   &   &   & 3889.6 & 46.3  &      & 6.32 &      & H8 3889.1 \\
  &    &   &   &   &         &        &      &      &      & He\,{\sc i} (m2) 3888.7 \\
  &    &   &   &   & 3934.1 & 30.5  & 0.11 & 0.80 & 0.07 & Ca\,{\sc ii} (m1) 3933.7 \\
3970.3 & 15.1 & &  6.93 & & 3970.7 & 45.3 & & 7.84 & & H$\epsilon$ 3970.1 \\
  &    &   &   &   &  &   &  &  &  & Ca\,{\sc ii} (m1) 3968.5 \\
  &    &   &   &   & 4003.0 & 30.0  &      & 0.02 &      & Fe\,{\sc ii} (m190) 4002.6 \\
  &    &   &   &   & 4009.6 & 22.5 & & 0.07 & & He\,{\sc i} (m55) 4009.3 \\
4026.8 & 44.7 &  & 0.41 & & 4026.5 & 22.4 & & 0.36 & & He\,{\sc i} (m18) 4026.2 \\
4102.1 & 29.3 & & 5.93 & & 4102.5 & 58.5 & & 6.82 & &  H$\delta$ 4101.7 \\
4121.6 & 43.7 & & 0.06 & & 4121.5 & 36.4 & & 0.02 & & He\,{\sc i} (m16) 4121.0 \\
4128.8 & 47.4 & & 0.17 & & 4128.6 & 36.3 & & 0.13 & & Si\,{\sc ii} (m3) 4128.1 \\
4131.6 & 47.2 & & 0.24 & & 4131.4 & 36.3 & & 0.14 & & Si\,{\sc ii} (m3) 4130.9 \\
4144.8 & 72.4 &  & 0.18 & & 4144.6 & 57.9 & & 0.14 & & He\,{\sc i} (m53) 4143.8 \\
        &    &   &   &   &  4164.1 &   &  & 0.02 & & Uid \\  
4173.6 & 7.2 &  &  0.06 &  & 4174.1 & 43.1 & & 0.08 &    & Fe\,{\sc ii} (m27) 4173.5 \\
4178.9 & 0.0 &  & 0.08 &  &  4179.4 & 36.0 & & 0.12 & & Fe\,{\sc ii} (m28) 4178.9 \\
        &      &  &      &  &  & & & & & Fe\,{\sc ii} (m21) 4177.7 \\
4233.7 & 35.4 & 0.06 & 0.09 &  0.11 & 4233.6 & 28.4 & 0.13 & 0.16 & 0.13 & Fe\,{\sc ii} (m27) 4233.2 \\
4244.5 & 35.3 &  &  & 0.04 & & & & & & [Fe\,{\sc ii}] (m21F) 4244.0 \\
        &    &   &   &   &  4258.8 & 42.3 & & 0.01 & & Fe\,{\sc ii} (m28) 4258.2 \\
        &    &   &   &   &  4262.6 & 49.3 & & 0.02 & & Cr\,{\sc ii} (m31) 4261.9 \\
        &      &  &      &  &  & & & & & Cr\,{\sc ii} (m17) 4261.8 \\
4268.0 &  70.3     &  & 0.04 & & 4267.8 &  56.3     & & 0.05 & & C\,{\sc ii} (m6) 4267.0 \\
        &      &  &      &  &  & & & & & C\,{\sc ii} (m6) 4267.2 \\
        &      &  &      &  &  & & & & & C\,{\sc ii} (m6) 4267.3 \\
       &    &   &   &   &   4273.9 & & & 0.01 & & Uid \\  
4277.3 & 35.1 & & & 0.03 & 4277.3 & 35.1 & & & 0.03 & [Fe\,{\sc ii}] (m21F) 4276.8 \\
4287.9 & 35.0 & & & 0.04 & 4287.9 & 35.0 & & & 0.08 & [Fe\,{\sc ii}] (m7F) 4287.4 \\
4290.3 & 7.0 & & 0.02 & & 4290.9 & 49.0 & & 0.04 & & Ti\,{\sc ii} (m41) 4290.2 \\
4294.2 & 7.0 & & 0.02 & & 4294.7 & 41.9 & & 0.04 & & Ti\,{\sc ii} (m20) 4294.1 \\
        &      & &      & & 4297.1 & 34.9 & & 0.06 & & Fe\,{\sc ii} (m28) 4296.6 \\
4300.5 & 27.9 & 0.06 & 0.02 & 0.10 & 4300.5 & 27.9 & & 0.07 & & Ti\,{\sc ii} (m41) 4300.1 \\
      &    &   &   &   &  4303.8 & 41.8 & & 0.04 & & Fe\,{\sc ii} (m27) 4303.2 \\
      &    &   &   &   &  4320.0 &  & & & 0.02 & Uid \\ 
4340.6 & 6.9 & & 4.30 & & 4340.9 & 27.7 & & 5.31 & & H$\gamma$ 4340.5 \\
4352.4 & 41.4 & & 0.07 & 0.07 & 4352.2 & 27.6 & 0.08 & 0.07  & 0.08 & Fe\,{\sc ii} (m27) 4351.8 \\
4359.6 & 20.7 & & & 0.09 & 4359.7 & 27.5 & & & 0.10 & [Fe\,{\sc ii}] (m7F) 4359.3 \\
        &       & & &      & &  &  &  &  & [Fe\,{\sc ii}] (m21F) 4358.4 \\
     &    &   &   &   &  4369.3 &  & & 0.06 &  & Uid \\ 
4385.4 & 54.7 & & 0.06 & & 4385.9 & 34.2 & & 0.13 & & Mg\,{\sc ii} 4384.6 \\
4388.6 & 47.9 &  & 0.15 & & 4388.6 & 47.7 & & 0.21 & & He\,{\sc i} (m51) 4387.9 \\
4391.4 & 54.7 &  & 0.07 & & 4391.2 & 41.0 & & 0.13 & & Mg\,{\sc ii} (m10) 4390.6 \\
        &       &  &      & &        &        & &      & & Fe\,{\sc ii} (m27) 4389.9 \\
4395.6 & 41.0 &  & 0.04 & 0.04 & 4395.5 & 34.1 & 0.10 & 0.02 & 0.08 & Ti\,{\sc ii} (m19) 4395.0 \\
4444.5 & 47.3 & & 0.02 & 0.04 & 4443.6 & -13.5 & 0.07 & 0.01 & 0.03 & Ti\,{\sc ii} (m19) 4443.8 \\
     &    &   &   &   &  4458.4 & 26.9 & &  & 0.03 & [Fe\,{\sc ii}] (m6F) 4458.0 \\
4472.4 & 47.0 &  & 0.34 & & 4472.2 & 33.5 & & 0.29 & & He\,{\sc i} (m14) 4471.7 \\
4482.2 & 60.3 &  &  0.47    & & 4482.0 & 43.4 & & 0.36 & & Mg\,{\sc ii} (m4) 4481.3 \\
     &    &   &   &   & 4489.8 & 40.1 & & 0.02 & & Fe\,{\sc ii} (m37) 4489.2 \\
      &    &   &   &   & 4492.1 & 46.8 & & 0.01 & & Fe\,{\sc ii} (m37) 4491.4 \\
4502.1 & 55.3 &  & 0.1  & 0.04 & 4501.1 & -11.3 & 0.07 & 0.01 & 0.03 & Ti\,{\sc ii} (m31) 4501.3  \\
4508.7 & 26.6 & & 0.04 & 0.02  & 4508.6 & 20.0 & 0.06 & 0.07 &  0.04 & Fe\,{\sc ii} (m38) 4508.3 \\
4515.8 & 39.9 & & 0.03 & 0.01 & 4516.0 & 46.5 & & 0.05 & & Fe\,{\sc ii} (m37) 4515.3 \\
        &       & & &  & & & & & & Fe\,{\sc ii} (m20) 4515.2 \\
4520.8 & 39.8 & & 0.02 & 0.02 & 4520.8 & 39.8 & & 0.01 & & Fe\,{\sc ii} (m37) 4520.2 \\
4523.2 & 39.8 & & 0.06 & 0.05 & 4523.3 & 46.4 & 0.04 & 0.10 & 0.06 & Fe\,{\sc ii} (m38) 4522.6 \\
4534.5 & 35.1 & & 0.03 & 0.04 & 4533.9 & -4.7 & 0.07 & 0.02 & 0.04 & Ti\,{\sc ii} (m50) 4534.0 \\
       &       & & &  & & & & & & Fe\,{\sc ii} (m37) 4534.2 \\
      &    &   &   &   & 4542.2 & 46.2 & & 0.01 & & Fe\,{\sc ii} (m38) 4541.5 \\
4550.0 & 33.0 & & 0.13 & 0.11 & 4549.9 & 26.4 & 0.14 & 0.16 & 0.12 & Fe\,{\sc ii} (m38) 4549.5 \\
4556.6 & 46.1 & & 0.04 & 0.04 & 4556.6 & 46.1 & 0.02 & 0.06 & 0.04 & Fe\,{\sc ii} (m37) 4555.9 \\
4558.9 & 26.3 & & 0.05 & 0.02 & 4559.4 & 52.7 & & 0.09 & & Fe\,{\sc ii} (m20) 4558.6 \\
4564.2 & 26.3 & & 0.01 & 0.02  & 4564.4 & 39.4 & & 0.03 & & Ti\,{\sc ii} (m50) 4563.8 \\
4572.5 & 32.8 & & 0.03 & 0.04 & 4572.5 & 32.8 & 0.09 & 0.01 & 0.07 & T\,{\sc ii} (m82) 4572.0 \\
      &    &   &   &   &  4576.8 & 32.8 & & 0.06 & & Fe\,{\sc ii} (m38) 4576.3 \\
4584.5 & 45.8 & & 0.11 & 0.13 & 4584.4 & 39.3 & 0.20  & 0.12 & 0.23 & Fe\,{\sc ii} (m38) 4583.8 \\
        &       & & &  & & & & & & Fe\,{\sc ii} (m26) 4584.0 \\
4588.3 & 6.5 & & 0.05 & & 4588.9 & 45.8 & & 0.06 & & Cr\,{\sc ii} (m44) 4588.2 \\
        &       & & &  & & & & & & Cr\,{\sc ii} (m16) 4588.4 \\
      &    &   &   &   & 4592.7 & 39.2 & & 0.02 & & Cr\,{\sc ii} (m44) 4592.1 \\
      &    &   &   &   & 4596.6 & 58.8 & & 0.04 & & Fe\,{\sc ii} (m38) 4595.7 \\
       &    &   &   &   & 4619.5 & 45.5  & & 0.03 & & Cr\,{\sc ii} (m16) 4618.8 \\
      &    &   &   &   &  &  & & & & C\,{\sc ii} 4618.9 \\
4630.2 & 58.3 & & 0.01 & 0.05 & 4630.1 & 51.8 & 0.07 & 0.04 & 0.07 & Fe\,{\sc ii} (m37) 4629.3 \\
       &    &   &   &   & 4635.5 & 58.3 & & 0.07 & & Fe\,{\sc ii} (m25) 4634.6 \\
        &    &   &   &   & 4640.1 & & & &  0.01 & Uid \\
4713.9 & 31.8 &  & 0.08 & & 4713.8 & 25.5 & & 0.06 & & He\,{\sc i} (m12) 4713.4 \\
       &    &   &   &   & 4732.2  & 50.7 & & 0.02 & & Fe\,{\sc ii} (m43) 4731.4 \\
4814.9 & 18.7 & &  & 0.05 & 4814.9 & 18.7 & &  &  0.05 & [Fe\,{\sc ii}] (m20F) 4814.6 \\
4824.4 & 18.7 & & 0.03 & & 4824.9 & 24.9 & 0.03 & 0.04 & 0.01  & Cr\,{\sc ii} (m30) 4824.1 \\
       &       & &      & & & & & & & S\,{\sc ii} (m52) 4824.1 \\
4848.5 & 18.6 & & & 0.03 & 4849.0 & 49.5 & & 0.04 & & Cr\,{\sc ii} (m30) 4848.2 \\
      &       & &      & & & & & & & Fe\,{\sc ii} (m30) 4847.6  \\
4862.8 & 92.6 & & 0.26 & & 4862.8 & 92.6 & & 0.21 & &  H$\beta$ 4861.3\footnote{Because of the complexity of the profile, the equivalent width of H$\beta$ (and also of H$\gamma$ and H$\delta$) was derived considering the total one, i.e., the equivalent width of the absorption component minus that one from the emission components.} \\
4876.7 & 18.5  &   & 0.02 &   &  4877.1 & 43.1 & & 0.03 & & Cr\,{\sc ii} (m30) 4876.4 \\
      &       & &      & & & & & & &  Cr\,{\sc ii} (m30) 4876.5 \\
4890.3 & 43.0   &   &   & 0.04  & 4890.1 & 30.7 & & & 0.03 & [Fe\,{\sc ii}] (m4F) 4889.6 \\
      &       & &      & & & & & & & [Fe\,{\sc ii}] (m3F) 4889.7 \\
      &    &   &   &   & 4905.9 & 30.6 & & & 0.02 & [Fe\,{\sc ii}] (m20F) 4905.4 \\
4924.8 & 54.8 & & 0.26 & 0.40 & 4925.0 & 67.0 & 0.20 & 0.29 & 0.35 & Fe\,{\sc ii} (m42) 4923.9 \\
5019.5 & 65.8 & 0.03 & 0.28 & 0.57 & 5019.4 & 59.8 & 0.31 & 0.34 & 0.48 & Fe\,{\sc ii} (m42) 5018.4 \\
5041.9 & 47.6 &  & 0.11 & & 5041.7 & 35.7 & & 0.08 & & Si\,{\sc ii} (m5) 5041.1 \\
5048.4 & 41.6 &  & 0.06 & & 5048.4 & 41.6 & & 0.03 & & He\,{\sc i} (m47) 5047.7 \\
5057.2 & 47.5 &  & 0.20 & & 5056.7 & 17.8 & & 0.12 & & Si\,{\sc ii} (m5) 5056.4 \\
         &    &   &   &   &  5094.1 & & & 0.01 & & Uid \\
        &    &   &   &   & 5101.6 & 35.3 & & 0.03 & & Fe\,{\sc ii} 5101.0 \\
5159.3 & 75.6 & & & 0.10 & 5159.1 & 64.0 & & & 0.13 & [Fe\,{\sc ii}] (m18F) 5158.0 \\
5169.7 & 40.6 & 0.13 & 0.42 & 0.48 & 5170.0 & 58.0 & 0.30 & 0.50 & 0.45 & Fe\,{\sc ii} (m42) 5169.0 \\
5198.7 & 63.5 & & 0.04 & 0.07 & 5198.3 & 40.4 & 0.10 & 0.05 & 0.11 & Fe\,{\sc ii} (m49) 5197.6 \\
        &    &   &   &   & 5228.0 &  & & 0.03 & & Uid \\ 
5235.6 & 57.3 & & 0.05 & 0.09 & 5235.2 & 34.4 & 0.09 & 0.06 & 0.06 & Fe\,{\sc ii} (m49) 5234.6 \\
5262.4 & 45.6 & & & 0.08 & 5262.3 & 39.9 & & & 0.06 & [Fe\,{\sc ii}] (m19F) 5261.6 \\
5276.6 & 34.1 & 0.07 & 0.03 & 0.12 & 5276.2  & 11.4 & 0.14 & 0.20 & 0.10 & Fe\,{\sc ii} (m49) 5276.0 \\
        &       &      &      &      &  5284.3 & 11.4 & 0.03 & 0.01 & 0.02 & Fe\,{\sc ii} (m41) 5284.1 \\
5317.8 & 67.7 & 0.06 & 0.02 & 0.30 & 5317.3 & 39.5 & 0.23 & 0.09 & 0.21 & Fe\,{\sc ii} (m49) 5316.6 \\
        &       &      &      &      & 5334.1 & 22.5 & & & 0.06 & [Fe\,{\sc ii}] (m19F) 5333.7 \\
5364.2 & 72.7 & & & 0.06 & 5363.1 & 56.0 & 0.06 & 0.03 & 0.05 & [Fe\,{\sc ii}] (m17F) 5362.1 \\
5377.0 & 27.9 & & & 0.02 & 5377.0 & 27.9 & & & 0.03 & [Fe\,{\sc ii}] (m19F) 5376.5 \\
       &       &      &      &      &  5467.7 &  & & 0.04 & & Uid \\
       &       &      &      &      &  5528.0 & 38.0 & & & 0.03 & [Fe\,{\sc ii}] (m17F) 5527.3 \\
       &       &      &      &      & & & & & & [Fe\,{\sc ii}] (m34F) 5527.6 \\
      &       &      &      &      & 5535.5 & 32.5 & 0.05 & 0.01 & 0.05 & Fe\,{\sc ii} (m55) 5534.9 \\
5577.9 & 32.3 & & & 0.04 & 5578.1 & 43.0 & & & 0.03 & [O\,{\sc i}] (m3) 5577.3 \\
5876.8 & 61.3 & 0.13 & 0.53 & & 5876.1 & 25.5 & 0.08 & 0.23 & 0.05 & He\,{\sc i} (m11) 5875.6 \\
5890.4 & 20.4 & & 0.73 & 0.19 & 5890.6 & 30.6 & & 0.59 & & Na\,{\sc i} (m1) 5890.0 \\
5896.5 & 30.5 & & 0.49 & 0.14 & 5896.6 & 35.6 & & 0.39 & & Na\,{\sc i} (m1) 5895.9 \\
      &       &      &      &      & 6149.3 & 4.9 & & 0.08 & & Fe\,{\sc ii} (m74) 6149.2 \\
6158.4 & 9.7 &  & 0.14 & & 6158.2 & 68.2 & & 0.12 & & O\,{\sc i} (m10) 6158.2 \\
        &       &  &      & &        &       & &      & & O\,{\sc i} (m10) 6156.8 \\
       &       &      &      &      & 6176.1 & 45.7 & & 0.03 & & Fe\,{\sc ii} (m200) 6175.2 \\ 
6238.6 & 9.6 & & 0.03 & & 6239.4  & 48.1 & & 0.08 & & Fe\,{\sc ii} (m74) 6238.4 \\
6248.6 & 48.0 & & 0.05 & 0.06 & 6248.7 & 52.8 & 0.03 & 0.05 & 0.06 & Fe\,{\sc ii} (m74) 6247.6 \\
6301.1 & 38.1 &  &  & 0.55 &  6301.1 & 38.1 & & & 0.58 & [O\,{\sc i}] (m1F) 6300.3 \\
6319.0 & 38.0 &  &  & 0.12 & 6318.9 & 33.2 & & & 0.19 & Mg\,{\sc i} (m23) 6318.2 \\
       &       &  &  &      & & & & & & Mg\,{\sc i} (m23) 6318.8 \\
       &       &  &  &      &  & & & & & Fe\,{\sc ii} 6318.0 \\
6348.6 & 70.9 &  & 0.54 & & 6347.5 & 18.9 & 0.05 & 0.32 & & Si\,{\sc ii} (m2) 6347.1 \\
6364.6 & 37.7 & & & 0.20 & 6364.6 & 37.7 & & & 0.21 & [O\,{\sc i}] (m1F) 6363.8 \\
6372.9 & 70.6 & & 0.35 & & 6371.7 & 14.1 & 0.04 & 0.22 &  & Si\,{\sc ii} (m2) 6371.4 \\
6385.6 &  86.9 & &  & 0.11 & 6385.1 & 63.5 & & & 0.18 & Fe\,{\sc ii} 6383.8 \\
       &       &  &  &      &  & & & & & Fe\,{\sc ii} (m203) 6387.6 \\
      &       &      &      &      &  6417.8 & 42.1 & & 0.04 & & Fe\,{\sc ii} (m74) 6416.9 \\
6457.7 & 60.4 & & 0.06 & 0.06 & 6457.1 & 51.1 & 0.10 & 0.07 & 0.10 & Fe\,{\sc ii} (m74) 6456.4 \\
6564.8 & 91.4 &  & & 61.35 & 6564.1 & 59.4 & & & 57.56 &  H$\alpha$ 6562.8 \\
6680.1 & 85.4 & & 0.34 & & 6678.7 & 22.5 & 0.03 & 0.12 & 0.03 & He\,{\sc i} (m46) 6678.2 \\
7067.0 & 76.4 & & 0.10 & & 7066.1  & 38.2 & 0.02 & 0.06 & 0.04 & He\,{\sc i} (m10) 7065.2 \\
       &       & &  & & &  & & & & He\,{\sc i} (m10) 7065.7 \\
7496.5 &    &  &  & 0.13  & 7497.1 & & & & 0.15 & Uid   \\
7515.0 &       & & & 0.31 & 7514.2 & & & & 0.36 & Uid \\
7713.0 & 50.6 & & & 0.26 & 7712.2 & 54.5 & 0.11 & 0.01 & 0.06 & Fe\,{\sc ii} (m73) 7711.7 \\
7773.5 & 57.9 & 0.20 & 1.61 & & 7774.4 & 92.6 & 0.25 & 1.77 & 0.41 & O\,{\sc i} (m1) 7772.0 \\
       &       & &  & & & & & & & O\,{\sc i} (m1) 7774.2 \\
       &       & &  & & & & & & & O\,{\sc i} (m1) 7775.4 \\
8325.3 &       & &  & 0.20 & & & & & & Uid \\
8346.9 & 32.4 & &  & 0.31 & & & & & & H\,{\sc i} P23 8346.0 \\
8360.0 & 35.9 & &  & 0.31 & & & & & & H\,{\sc i} P22 8359.0 \\
8375.5 & 35.8 & &  & 0.61 & 8374.8 & 10.8 & & & 0.86 & H\,{\sc i} P21 8374.5 \\
8393.5 & 39.3 & &  & 0.80 & 8392.4 & 0.0 & & & 1.07 & H\,{\sc i} P20 8392.4 \\
8414.2 & 32.1 & &  & 0.75 & 8413.7 & 14.3 & &  & 1.33  & H\,{\sc i} P19 8413.3 \\
8440.0 &       & &  & 0.76 & 8435.8 & & & & 0.85 &  Uid \\
8448.1 & 46.2 & & & 4.01 & 8447.1 & 10.7 & & & 3.70 & O\,{\sc i} (m4) 8446.8 \\ 
8468.8 & 53.2 & & & 1.08 & 8468.6 & 46.1 & & & 1.40 & H\,{\sc i} P17 8467.3 \\
8501.4 & -38.8 & & & 2.35 & 8500.9 & 0.0 & & & 2.29 & H\,{\sc i} P16 8502.5 \\
       &       &  &  &      & & & & & & Ca\,{\sc ii} (m2) 8498.0 \\
8598.8 & 14.0 & & & 1.85 & 8598.2 & -7.0 & & & 1.86 &  H\,{\sc i} P14 8598.4 \\
8630.7 & 52.2 & & & 1.14 & 8630.6 & 48.7 &   & & 1.73 & N\,{\sc i} (m8) 8629.2 \\
8664.9 & -3.5 & & & 2.79 & 8664.3 & -24.2 & & & 2.94 & H\,{\sc i} P13 8665.0 \\
       &       &  &  &      & & & & & & Ca\,{\sc ii} (m2) 8662.1 \\
8684.0 & 20.7 & & & 0.47 & & & & & & N\,{\sc i} (m1) 8683.4 \\
8751.8 & 44.6 & & & 2.24  & 8751.9 & 48.0 & & & 2.27 & H\,{\sc i} P12 8750.5 \\
\label{identification}
\end{longtable}
}

\begin{itemize}

\item Hydrogen

\end{itemize}

\begin{figure}[t!] 
\resizebox{\hsize}{!}{\includegraphics{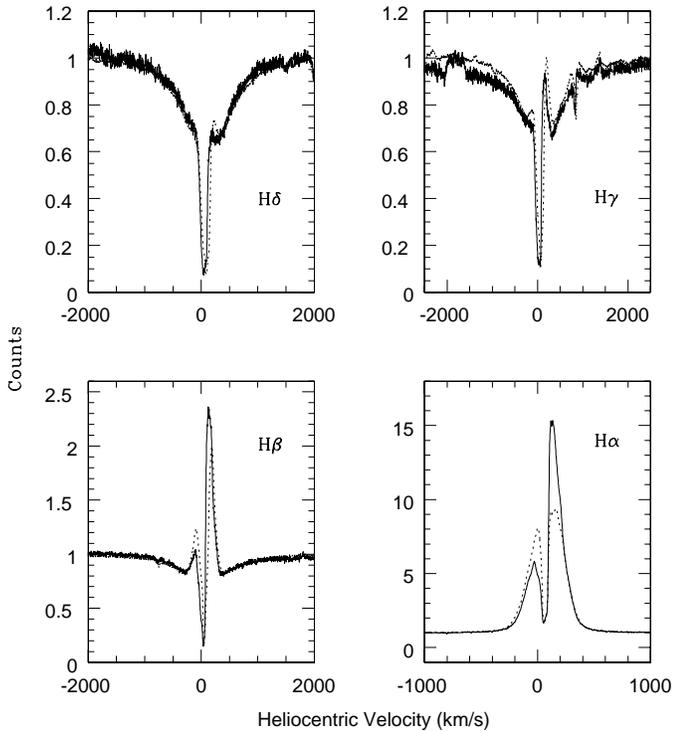}}
\caption{Balmer lines seen in the FEROS spectra of HD\,50138. The solid line is related to the spectrum taken in 1999 and the dotted line to the 2007 one.}
\label{Hidro}
\end{figure}

In our FEROS spectra, we identify the presence of Balmer lines (see Table\,\ref{identification}). Four of them, H$\delta$, H$\gamma$, H$\beta$, and H$\alpha$ can be seen in Fig.\,\ref{Hidro}. The profiles of these lines, except for H$\alpha$, are composed of a broad absorption component, probably of photospheric origin, and an emission and a narrow absorption components superimposed, which are probably formed in the circumstellar medium. This kind of profile was already described by Houziaux (\cite{Houziaux}) from observations taken in the late fifties. The H$\alpha$ line presents a double-peaked structure, where the V/R changes from 0.36 in 1999 to 0.88 in 2007. These values are different than those found by Halbedel (\cite{Halbedel}) and Pogodin (\cite{Pogodin}) of 0.6 and 0.5, respectively. However, the peak separation presents small differences in both our spectra, $\sim$ 160 km\,s$^{-1}$ in 1999 and 150 km\,s$^{-1}$ in 2007, in agreement with the value cited by Oudmaijer \& Drew (\cite{Oudmaijer}) of $\sim$ 160 km\,s$^{-1}$. 

\begin{figure}[t!] 
\resizebox{\hsize}{!}{\includegraphics{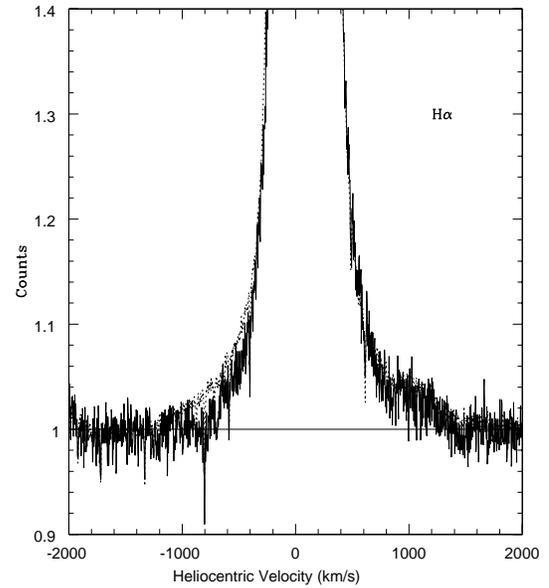}}
\caption{H$\alpha$ wings seen in the FEROS spectra of HD\,50138. The different line styles have the same meaning as in Fig. 1. The normalized continuum is also presented for better visualization.}
\label{Halpha}
\end{figure}

Another significant variation is related to the radial velocities of the narrow absorption components. In 1999, this value is almost constant for H$\delta$, H$\gamma$, and H$\beta$, $\sim$ 35 km\,s$^{-1}$. However in 2007, this value is quite different for each line, 95 km\,s$^{-1}$ (H$\delta$), 60 km\,s$^{-1}$ (H$\gamma$) and 50 km\,s$^{-1}$ (H$\beta$). The H$\alpha$ line's central absorption presents a constant value of $\sim$ 60 km\,s$^{-1}$ on both dates. This velocity is quite different from the velocity reported by Oudmaijer \& Drew (\cite{Oudmaijer}); however, this can be related to different spectral resolutions. As cited by Pogodin (\cite{Pogodin}), the Balmer lines also present symmetric wings, which show velocities that can reach $\sim$ 1500 km\,s$^{-1}$, as can be seen in Fig.\,\ref{Hidro} and even better in the blow-up of the H$\alpha$ wings in Fig.\,\ref{Halpha}. However, these high velocities seem to come from electron scattering, since no other element identified in our data present such characteristic.

\begin{figure}[t!] 
\resizebox{\hsize}{!}{\includegraphics{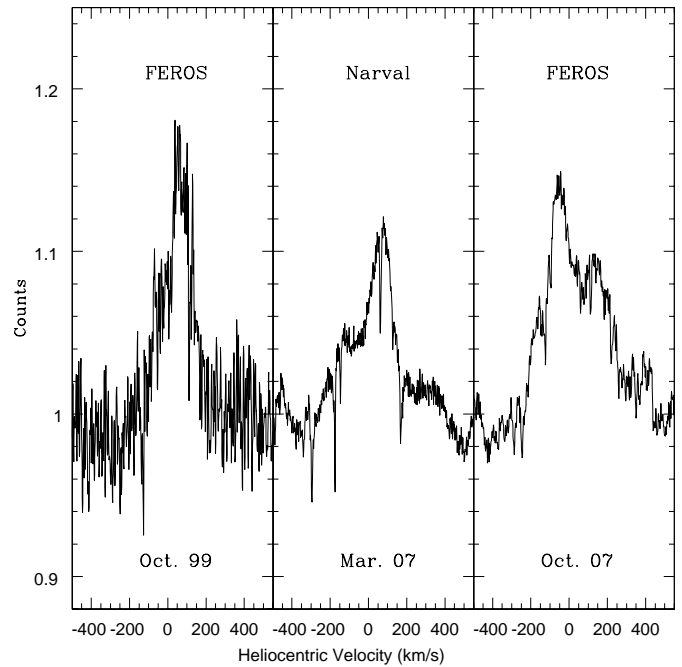}}
\caption{Variations in the line profile of Paschen 19 at 8413\AA.}
\label{Paschen}
\end{figure}

We identify Paschen 22 to Paschen 12 lines in our spectra. We show one of these lines, Paschen 19, in Fig.\,\ref{Paschen}. They display strong variations in our FEROS and Narval data. In the 1999 FEROS data, we detect a single-peaked profile with a blue shoulder. This shoulder becomes more prominent, probably because of the better S/N, in the Narval data. However, with the 2007 FEROS data, we clearly have a double-peaked profile with a peak separation of 190 km\,s$^{-1}$. Assuming the shoulder previously seen in 1999 and in the Narval data as another peak, we have a variation of V/R in the 2007 FEROS data. A similar variation was reported by Andrillat \& Houziaux (\cite{Andrillat}), which was associated to an outburst. The wings of Paschen lines are extended in all spectra to $\sim$ 300 km\,s$^{-1}$. However, as seen in Table\ref{identification}, we believe the radial velocities obtained for the Paschen lines are affected by some problems with the precision of the wavelength calibration of FEROS, beyond 8300 \AA.

\begin{itemize}

\item Iron

\end{itemize}

Iron, as can be seen in Table\,\ref{identification}, is by far the element with the most of lines identified in our spectra, with all of them from Fe{\sc ii}. We have identified permitted and forbidden lines. The latter are represented by narrow single-peaked emission lines from multiplets 3, 4, 6, 7, 17, 18, 19, 20, 21, and 34. These forbidden lines do not present any noticeable difference in all spectra. 

\begin{figure}[t!] 
\resizebox{\hsize}{!}{\includegraphics{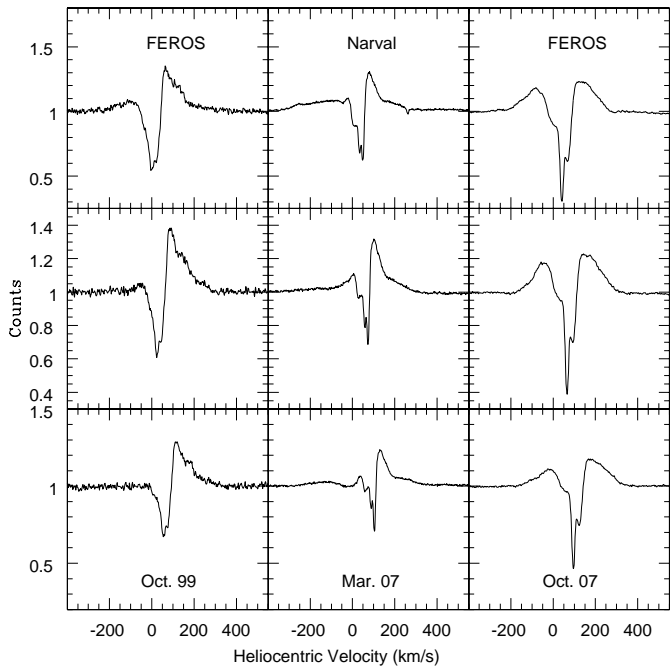}}
\caption{Variations in the line profiles of the Fe{\sc ii} multiplet 42. The bottom line corresponds to the line at 4923$\AA$, the middle one at 5018$\AA$, and the top one at 5169$\AA$. }
\label{Iron}
\end{figure}

We have identified permitted lines from the multiplets 20, 21, 25, 26, 27, 28, 30, 37, 38, 41, 42, 43, 49, 55, 73, 74, 167, 190, 200, and 203. Unlike Jaschek \& Andrillat (\cite{Jaschek1}), where the Fe{\sc ii} lines show absorption (mainly) or emission profiles, our spectrum presents a special kind of profile for each date (see Fig.\,\ref{Iron}). In the 1999 FEROS spectrum, these lines have a P-Cygni profile, where the absorption component is made up of two components, with the bluer one being more intense. In the 2007 FEROS spectrum, these lines have a different profile. There is an emission profile associated to a strong central absorption that also has two components, where the bluer one is also more intense. These absorption components are shifted to higher velocities compared to 1999. On the other hand, the Narval data show a sort of transition between the P-Cygni profile and the profile seen in the 2007 FEROS spectrum. The absorption components are not as intense as seen in the other spectra; however, three of them can be seen, where the red one, contrary to the FEROS data, is the most intense.

\begin{itemize}

\item Oxygen

\end{itemize}

Permitted and forbidden lines of neutral oxygen are seen in the HD\,50138 spectra. No O{\sc ii} lines were identified, indicating that the emission region is not highly ionized or that the higher ionized regions are rather small in volume, supressing the efficient production of observable O{\sc ii} emission and absorption. The forbidden lines present narrow emission profiles and do not show any significant differences in all the spectra, with an almost constant radial velocity around 40 km s$^{-1}$ (Table\,\ref{identification}). All three lines from the multiplet 1 are seen, different than reported by Andrillat \& Houziaux (\cite{Andrillat2}). In addition, a contamination by airglows, based on the comparison with spectra from the sky taken at the same time, is discarded. 

\begin{figure}[t!] 
\resizebox{\hsize}{!}{\includegraphics{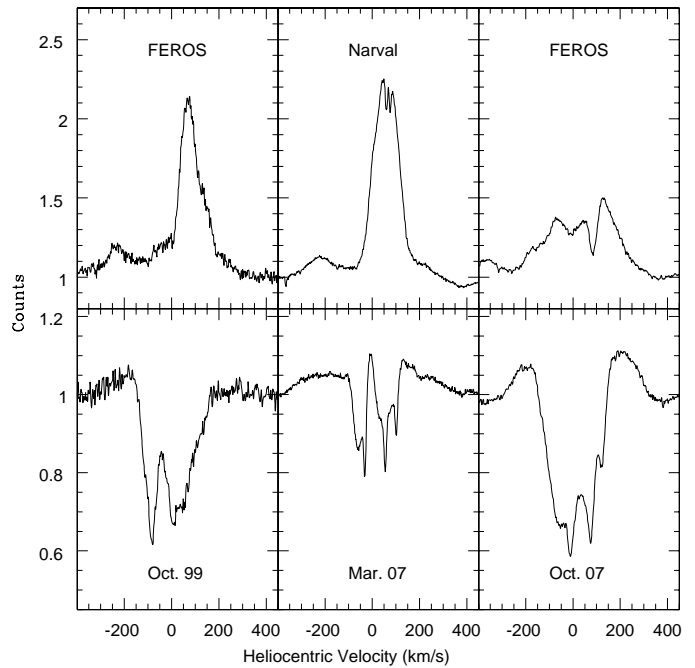}}
\caption{Profiles of the permitted lines of O{\sc i} and their variation. The bottom line corresponds to the triplet at 7774$\AA$ and the top one at 8446$\AA$.}
\label{Oxygen}
\end{figure}

The permitted lines present a complex profile and show a strong variation in the different spectra (see Fig.\,\ref{Oxygen}). The line centered at 7774\AA \ is in fact a triplet, thus the profile seen in our data has a combination of possible emission and absorption components from each line. However, the profile of this triplet changes strongly at different dates, as already mentioned by Jaschek \& Andrillat (\cite{Jaschek1}). Unlike the 1999 data, the 2007 FEROS data show the absorption components clearly superimposed over a strong emission component, similar to the Fe{\sc ii} lines. In the Narval spectrum, this line has a transitional profile between those detected in the different FEROS spectra, especially in the emission component.

Another oxygen line present in our spectra is located at 8446$\AA$. This line presents a strong emission in the FEROS spectrum of 1999 and in the Narval data; however, this profile changes completely in the FEROS spectrum of 2007, because it is less intense and has multiple peaks.

\begin{itemize}

\item Helium

\end{itemize}

\begin{figure}[t!] 
\resizebox{\hsize}{!}{\includegraphics{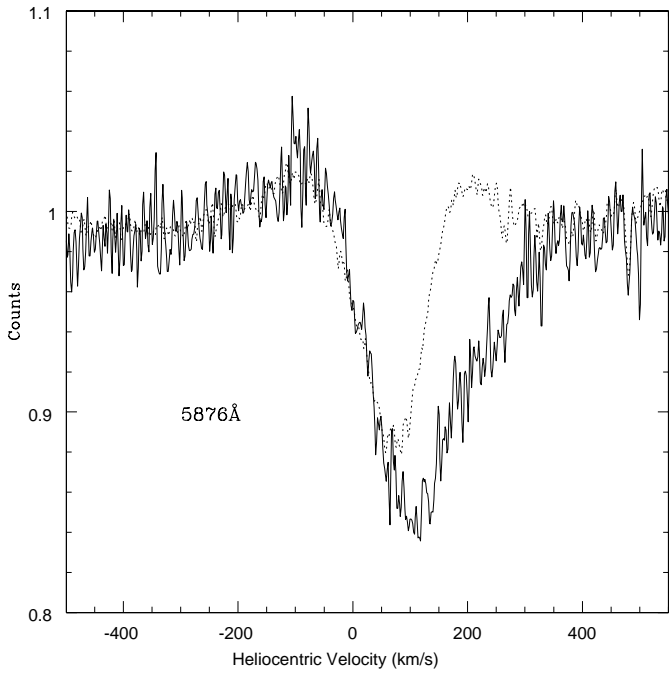}}
\caption{Line profile variation of the He{\sc i} line at 5876$\AA$. The solid line is related to the spectrum taken in 1999 and the dotted line to the 2007 one.}
\label{Helium}
\end{figure}

HD\,50138 presents several He{\sc i} lines from multiplets 2, 10, 11, 12, 14, 16, 18, 20, 22, 46, 47, 51, 53, and 55. Line profile variations are clearly seen in the different spectra (Fig.\,\ref{Helium}). In 1999, we can see an inverse P-Cygni profile for some lines, such as 5876$\AA$. The separation seen between the emission and the absorption components agrees with the value cited by Bopp (\cite{Bopp}), $\sim$ 200 km s$^{-1}$. However, in 2007, profiles similar as seen for the Fe{\sc ii} lines are formed, where the absorption component is narrower than in 1999. Both profiles present wings with similar extension velocities, as can be seen in Fig.\,\ref{Helium}. Similar variations were described by Bopp (\cite{Bopp}), who reported an inverse P-Cygni profile, a few months after the outburst in 1991, and also by Pogodin (\cite{Pogodin}), who cited that the profile of the line at 5876$\AA$, changes from night to night.

Our spectra do not show any He{\sc ii} line. The tentative detection of the He{\sc ii} line at 10123\AA, as reported by Jaschek \& Andrillat (\cite{Jaschek1}), could not be confirmed by our Narval data. We can therefore claim that this object is not very hot.

\begin{itemize}

\item Magnesium

\end{itemize}

\begin{figure}[t!]
\resizebox{\hsize}{!}{\includegraphics{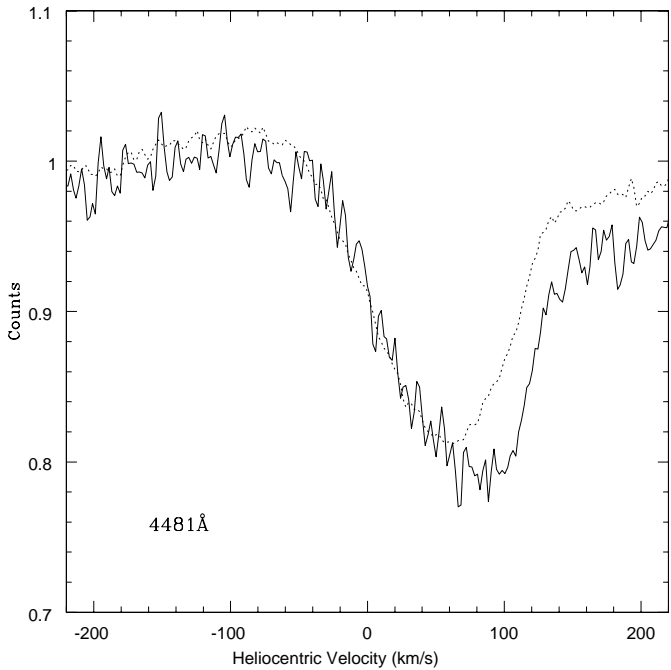}} 
\caption{As in Fig.\,\ref{Helium} but for the Mg{\sc ii} line at 4481$\AA$.}
\label{Magnesium}
\end{figure}

Only one line from Mg{\sc i}, at 6318$\AA$, is identified in our spectra. On the other hand, few lines from Mg{\sc ii} are seen in our data, with the most prominent one, at 4481$\AA$. This line mainly has photospheric origin, and as can be seen in Fig.\,\ref{Magnesium}, it also presents a noticeable variation in our FEROS data taken with 8 years of difference.

\begin{itemize}

\item Silicon

\end{itemize}

\begin{figure}[t!]
\resizebox{\hsize}{!}{\includegraphics{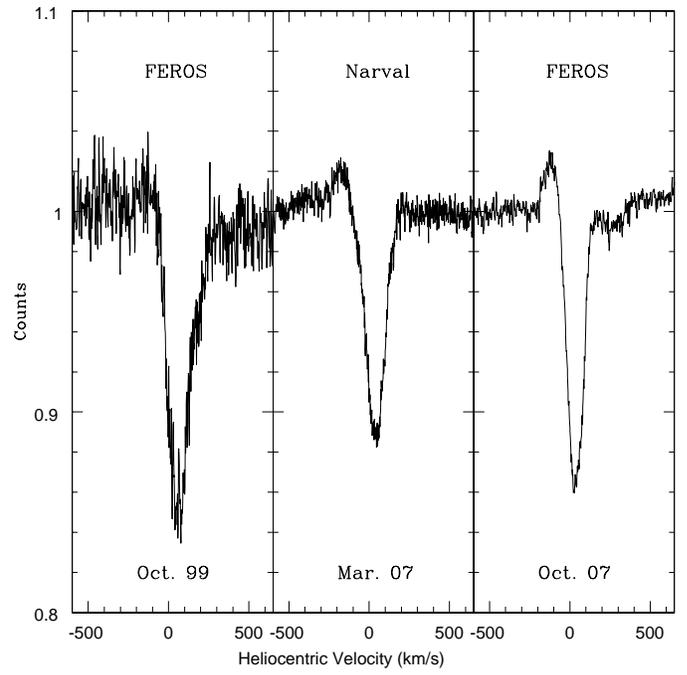}}
\caption{The variation of the line profile of Si{\sc ii} at 6347$\AA$.}
\label{Silicon}
\end{figure}

\begin{figure}[t!]
\resizebox{\hsize}{!}{\includegraphics{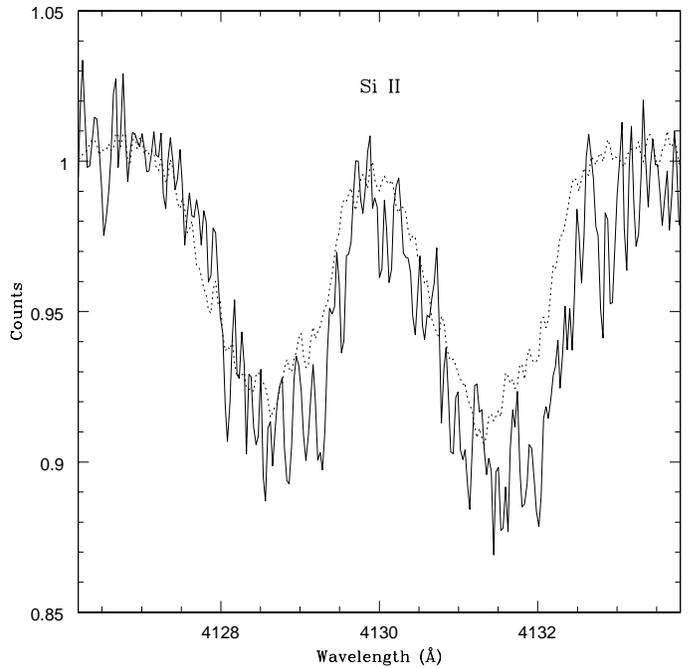}} 
\caption{As in Fig.\,\ref{Helium} but for the Si{\sc ii} lines at 4128$\AA$ and 4131$\AA$.} 
\label{Si_ii}
\end{figure}

In the FEROS spectrum of 1999, Si{\sc ii} lines have pure absorption profiles. On the other hand, in March 2007, the lines of the multiplet 2 ($\lambda$$\lambda$ 6347,6371) developed into an inverse P-Cygni profile, which is even more pronounced in October 2007, as can be seen in Fig.\,\ref{Silicon} for one of these lines. However, the lines of the multiplet 3 ($\lambda$$\lambda$ 4128,4131) still show pure absorption profiles at these dates (see Fig.\,\ref{Si_ii}) and might thus be considered as having a pure photospheric origin.

\begin{itemize}

\item Sodium

\end{itemize}

\begin{figure}[t!] 
\resizebox{\hsize}{!}{\includegraphics{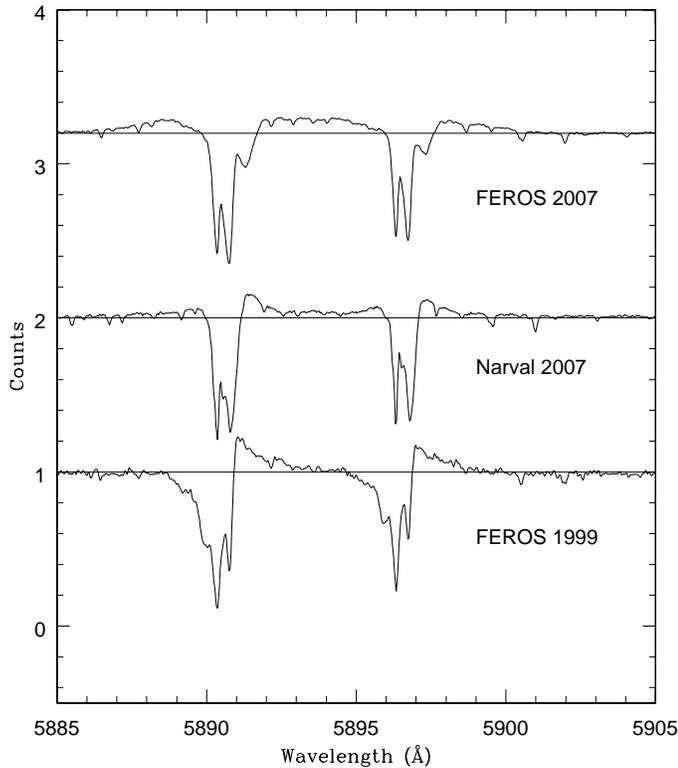}}
\caption{Different line profiles of the Na{\sc i} lines at 5890$\AA$ and 5896$\AA$. The normalized continuum is also presented for each date. }
\label{Sodium}
\end{figure}

Lines of Na{\sc i} are present in our spectra at 5890$\AA$ and 5895$\AA$. These lines are formed by a combination of circumstellar and interstellar contributions. However, it is impossible for us to disentangle each component, which makes it impossible to derive a distance to HD\,50138 from these lines. These lines also present variability from 1999 to 2007 (Fig.\,\ref{Sodium}), because it is almost a P-Cygni profile, with three absorption components in 1999 and a broad emission component superimposed by three absorption components in 2007. 

\begin{itemize}

\item Other lines

\end{itemize}

Permitted lines from Ti{\sc ii}, Ca{\sc ii}, Cr{\sc ii}, C{\sc ii}, S{\sc ii}, and N{\sc i} are also identified in our spectra (see Table\,\ref{identification}). These lines also show significant differences from 1999 to 2007, following the same behavior as described for the Fe{\sc ii} lines. 

\subsection{Determination of the extinction and spectral classification}\label{SpType}

The determination of the stellar parameters of HD\,50138 is difficult, since both its spectral lines and its continuum show strong variations. It is,
therefore, not surprising that the range of spectral types (from earlier
than B5 down to A0) and luminosity classes (I-V) found in the literature is
rather wide, because we can expect that these determinations depend on the method (or lines) used for the classification, as well as on the period at which the observations were obtained.

\begin{figure}[t!]
\begin{center}
\resizebox{\hsize}{!}{\includegraphics{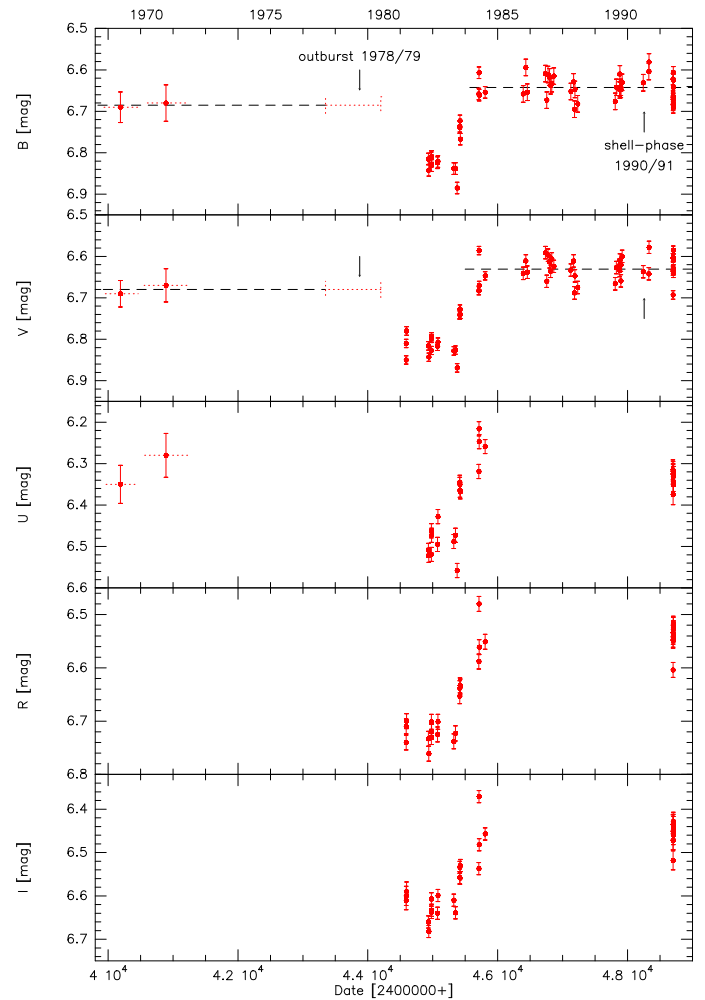}}
\caption{Time variation of photometric observations in the $UBVRI$ bands 
collected from the literature (see Table\,\ref{refs}). Also shown are the 
supposed positions of the outburst and the shell phase reported in the 
literature. The dashed lines in the $B$ and $V$ bands indicate the mean values 
before and after the outburst. The dotted bar in the $B$ and $V$ light-curves refer to the observations of 
Alvarez \& Schuster (\cite{Alvarez}).}
\label{ubvri}
\end{center}
\end{figure}

To constrain the spectral classification of HD\,50138, we made use of both our
own sets of high-resolution optical spectra and public photometric data. We collected all available $UBVRI$ band photometric data. These 
observations cover a time interval of almost 25 years and are displayed in 
Fig.\,\ref{ubvri}. The references for the data are listed in Table\,\ref{refs}.

\begin{table}[t!]
\caption{References to the $UBVRI$ observations of HD\,50138 shown in 
Fig.\,\ref{ubvri}.}
\label{refs}
\begin{center}
\begin{tabular}{ll}
    \hline
    \hline
 Julian Dates & Reference \\
 $[$2400000+] &  \\
    \hline
39916 -- 40464 & Haupt \& Schroll (\cite{Haupt}) \\
40556 -- 41224 & Penston (\cite{Penston}) \\
43350 -- 44200 & Alvarez \& Schuster (\cite{Alvarez}) \\
44589 -- 44596 & de Winter et al. (\cite{deWinter}) \\
44934 -- 45809 & Kilkenny et al. (\cite{Kilkenny}) \\
46392 -- 48331 & Halbedel (\cite{Halbedel}) \\
48699 -- 48707 & de Winter et al. (\cite{deWinter}) \\
    \hline
\end{tabular}
\end{center}
\end{table}

Unfortunately, hardly any photometric data could be found before and around the 
time of the outburst, which was reported to have happened in 1978/79
(Hutsem\'ekers \cite{Hutsemekers}). The consequences of this outburst are, however, clearly
visible in the light curves displayed in Fig.\,\ref{ubvri}: a fading of the
star in the $UBV$ bands by about 0.2 mag, and a recovery to the old brightness
after about 5 years. This trend in photometric data might be explained
by some large mass ejection during the outburst, hiding the stellar continuum
inside a dense shell or ring of (partly) optically thick material, which then turns transparent
during expansion, allowing the underlying star to be seen again. On the other hand, the shell phase seen in the optical spectra in 1990/91 
(Andrillat \& Houziaux 1991) remained invisible in the photometric data, as cited by Halbedel (1991). This might indicate that the
shell phase was caused by mass ejection with much less mass loss.

Interestingly, after the recovery of the star, it seems to be even brighter than 
it was before the outburst, as can be seen by the comparison of the mean values before and after the outburst (Fig.\,\ref{ubvri}). We also notice that the mean 
value before is based only on two datapoints, but even considering the large 
errorbars of these observations, it seems that the star appears brighter after 
$\sim 1984$. In addition, Alvarez \& Schuster (\cite{Alvarez}) observed HD\,50138 in a period close to the outburst, but without providing either the exact dates or the observed 
magnitudes. However, these authors claimed that during their observations, HD\,50138 
was not variable, compared to previous observations, meaning that $\Delta B$ and $\Delta V < 0.02$\,mag, according 
to their definition of variability.

This apparent brightening after $\sim 1984$ might be caused by some contribution 
of the circumstellar material ejected during the outburst and after adding flux at 
all bands. Even though HD\,50138 is known to possess a strong infrared excess 
emission due to circumstellar dust (Allen \cite{Allen}), this dust will hardly contribute at optical 
wavelengths, i.e., in the $UBV$ bands. On the other hand, based on 
spectropolarimetric observations performed in 1995 by Bjorkman et al. 
(\cite{Bjorkman}), HD\,50138 seems to have an almost edge-on ionized Be-like 
disk. Such a disk, which might have been formed from the outburst material in a 
similar way to how disks are formed around classical Be stars, can well add flux to the 
optical continuum in the form of free-free and especially free-bound emission. 

Whether the additional continuum emission from the ionized circumstellar
material/disk will hamper proper spectral classification of HD\,50138, or 
whether we even have to account for circumstellar dust along the line of sight
towards the star, acting as a further (i.e. circumstellar) extinction source,
will be discussed and investigated in detail in Sects.\,\ref{iondisk} and \ref{dust},
respectively.

\subsubsection{The interstellar extinction}\label{extinction}

\begin{figure}[t!]
\begin{center}
\resizebox{\hsize}{!}{\includegraphics{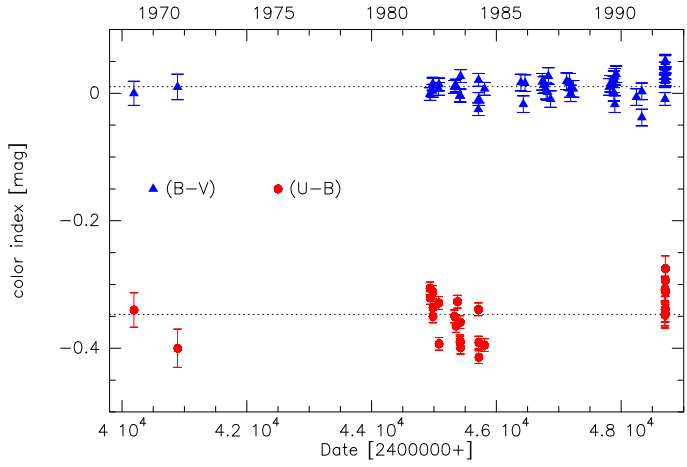}}
\caption{Time variation of the observed $(U-B)$ and $(B-V)$ color indices.}
\label{ub_bv}
\end{center}
\end{figure}

First, we estimate the total extinction towards HD\,50138, i.e., neglect any circumstellar contribution and thus derive an upper limit on the interstellar one. For this, we make use of the two color indices $(U-B)$ and $(B-V)$. Their 
variations with time are plotted in Fig.\,\ref{ub_bv}.

\begin{figure}[t!]
\begin{center}
\resizebox{\hsize}{!}{\includegraphics{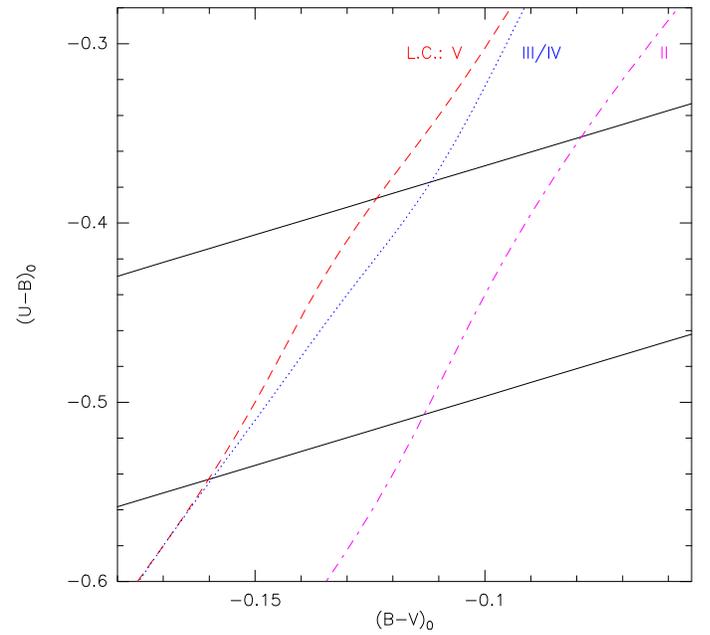}}
\caption{Constraining the range in spectral type for different luminosity
classes. The solid lines correspond to the upper and lower limits derived
from observations.}
\label{lc_det}
\end{center}
\end{figure}

Obviously, $(U-B)$ is quite variable, while $(B-V)$ remains rather stable, even 
during the outburst phase. Despite the variability in $(U-B)$, we can obtain a 
reasonable range for the possible stellar classification from these observed 
color indices, making use of the following relations for the interstellar extinction determinations, 
\begin{eqnarray}
E(B-V) & = & (B-V) - (B-V)_0 \\
E(B-V) & = & \frac{1}{0.77} E(U-B) = \frac{1}{0.77}\left\{(U-B)-(U-B)_0
\right\}\, .
\end{eqnarray}
The second relation follows from Leitherer \& Wolf (\cite{Leitherer}), and the 
parameters $(U-B)_0$ and $(B-V)_0$ refer to the intrinsic color indices. 
By combining these two equations, we obtain for each observed set of 
$(U-B)$ and $(B-V)$ colors a reddening independent equation of the form
\begin{equation}
(U-B)_0 = 0.77 \left\{(B-V)_0 - x\right\},
\label{UB_UV}
\end{equation}
relating the two intrinsic colors, with
\begin{equation}
x = (B-V) - \frac{(U-B)}{0.77}\, .
\end{equation}
We calculate the parameter $x$, hence the relation 
Eq.\,(\ref{UB_UV}), for each observation. Then we search for the 
two boundary cases, i.e., the highest and lowest values for $x$, and plot
the resulting range in relation Eq.\,(\ref{UB_UV}) in
Fig.\,\ref{lc_det}. Next, we look up the tables of Schmidt-Kaler 
(\cite{SK}) for the intrinsic colors of stars in the spectral
range B3 to A1. The luminosity class of HD\,50138 is not well known, but 
previous classifications found in the literature tend towards luminosity
classes III-V (e.g., Houziaux \cite{Houziaux}; Houziaux \& Andrillat
\cite {HouziauxAndrillat}; Fr\'{e}mat et al. \cite{Fremat}). Based on the Hipparcos distance, we 
can already exclude a supergiant classification for HD\,50138, since it 
would deliver an unrealistically high extinction value.
We thus restrict it to luminosity classes II to V. 

The behavior of the intrinsic colors from Schmidt-Kaler (\cite{SK}) 
is included in Fig.\,\ref{lc_det}. From the overlap with the observational 
ranges we find some preliminary sets, as listed in 
Table\,\ref{av_max}, of MK types, effective temperatures and color excess values, which are upper limits to the real interstellar extinction. The assignment of the temperature ranges is made based on
the tables of Flower (\cite{Flower}).

\begin{table}[t!]
\caption{Possible stellar classifications of HD\,50138 under the assumption 
of pure interstellar extinction along the line of sight.}
\label{av_max}
\begin{center}
\begin{tabular}{llcc}
    \hline
    \hline
 L.C. & Sp.Type & $T_{\rm eff} [K]$ & $E(B-V)_{\rm max} [mag]$\\
    \hline
 V     & B5.5--7.5 & 13\,300 $\pm$ 900    & 0.15 $\pm$ 0.03 \\
III/IV & B5.5--8.0 & 13\,100 $\pm$ 1\,100 & 0.15 $\pm$ 0.03 \\
 II    & B7.5--8.5 & 11\,600 $\pm$ 400    & 0.11 $\pm$ 0.02 \\
    \hline
\end{tabular}
\end{center}
\end{table}

A rather low value for the interstellar extinction has already been 
suggested by Houziaux \& Andrillat (\cite{HouziauxAndrillat}) and Hutsem\'ekers (\cite{Hutsemekers}) based on the weakness of the interstellar feature at $\lambda 2175\,\AA$ and on the absence of interstellar lines in the IUE spectrum. To split the total extinction value into its 
interstellar and (possible) circumstellar contributions we, therefore, search 
for further extinction indicators in our spectra. One such indicator makes use
of diffuse interstellar bands (DIBs). In our high-resolution spectra, we found 
one reasonably good DIB at $\lambda 5780\,\AA$, from which the extinction 
can be derived (see Herbig \cite{Herbig}). This feature is present in both the 
1999 and the 2007 FEROS spectra, but because of the poor S/N of the 1999 
data we can rely only on our 2007 data. From the equivalent width of 0.04\,\AA \ of 
this DIB, we derive an extinction value of $E(B-V) = 0.08\pm 0.01$\,mag. 

As a further check we investigate the interstellar extinction within the
galactic plane in the region around our target. It turns out that HD\,50138 is 
located within a region of rather low extinction (see Neckel \& Klare \cite{Neckel}; Chen et al. \cite{Chen}; 
Arenou et al. \cite{Arenou}). Using the model developed by Hakkila et al. (1997) for the large-scale visual interstellar extinction, assuming $R_{V} = A_{V}/E(B-V) = 3.1$, we have obtained $A_{V} = 0.22\pm 0.22$\,mag, delivering $E(B-V) = 0.07\pm 0.07$\,mag.
This value agrees with the value we found from the DIB. However, because of higher uncertainty (and definitely larger error) with this method, 
the value found from the DIB of $E(B-V) = 0.08\pm 0.01$\,mag seems to be reasonably 
accurate, leading us to use this value later. Compared to the (upper limit) values 
we found from the color indices, the contribution of the interstellar 
extinction towards HD\,50138 is thus only half of it, indicating that the other 
part must be circumstellar in nature.   

\subsubsection{The influence of circumstellar dust}\label{dust}

The presence of a strong infrared excess implies that HD\,50138 must be 
surrounded by circumstellar dust (Allen \cite{Allen}). How this dust is distributed, i.e., whether 
it is situated within a disk or shell, within or outside the line of sight,
is not known. 

The circumstellar dust, like the interstellar one, absorbs and scatters the light 
from the star, so that we can define a circumstellar extinction parameter, 
$A_{UBV}^{\rm dust}$. The intrinsic magnitude of the star in each photometric 
band, $UBV_{0}$, is then obtained from the observed one, corrected for the 
amounts of interstellar and circumstellar extinction, i.e,
\begin{equation}
UBV_{0} = UBV_{\rm obs} - \left(\frac{A_{UBV}}{A_{V}}\right)_{\rm ISM} 
A_{V}^{\rm ISM} - A_{UBV}^{\rm dust}\, ,
\label{ubv0dust}
\end{equation}
with the parameters $\left(A_{UBV}/A_{V}\right)_{\rm ISM}$ resulting from the 
mean interstellar extinction curve (see e.g. Cardelli et al. \cite{Cardelli}; 
Mathis \cite{Mathis}). A star that suffers from additional circumstellar 
extinction thus appears fainter.

The dust extinction parameter, $A_{UBV}^{\rm dust}$, can be calculated from
\begin{equation}
A_{UBV}^{\rm dust} = \frac{1}{0.4 \ln(10)}\frac{\sigma_{UBV}^{\rm ext}}
{\sigma_{V}^{\rm ext}} \tau_{V}^{\rm dust}\, ,
\end{equation}
with $\sigma^{\rm ext}$ as the extinction coefficient of the dust within the 
corresponding band and $\tau_{V}^{\rm dust}$ as the optical depth of the dust
at visual wavelengths. The extinction coefficients depend on the size of the
grains and their chemical composition. Since the dust composition along the
line of sight is not known, we allow for different sizes of the (spherically)
test grains, which we consider as composed either of amorphous carbon or
silicates. The range of sizes used extends from $7.5\times 10^{-3}\,\mu$m up to 
$\sim 1\,\mu$m. In addition to single grains of different sizes, we consider the well-known MRN grain size distribution (Mathis et al. \cite{MRN}).

\begin{figure}[t!]
\begin{center}
\resizebox{\hsize}{!}{\includegraphics{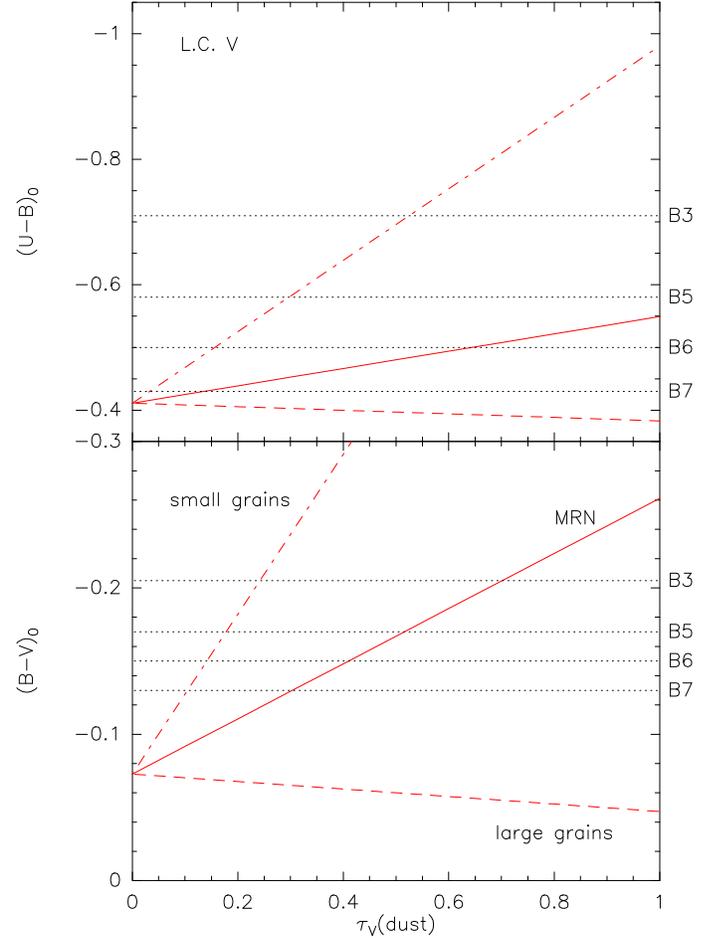}}
\caption{Intrinsic colors as a function of dust optical depth. Individual 
curves represent calculations performed for a mixture of silicates and amophous
carbon grains for either single (smallest and largest, as cited in the text) grain sizes or an MRN grain
size distribution. The dotted lines indicate theoretically expected intrinsic 
colors for different spectral types.}
\label{lc_dust}
\end{center}
\end{figure}

For each grain size and species, the dust extinction depends on only one free 
parameter, which is $\tau_{V}^{\rm dust}$. Expressing the set of
Eqs.\,(\ref{ubv0dust}) by the intrinsic color indices $(U-B)_0$ and $(B-V)_0$,
we end up with two equations 
\begin{eqnarray}
(U-B)_0 & = & f(\tau_{V}^{\rm dust})\,, \\
(B-V)_0 & = & f(\tau_{V}^{\rm dust})\,.
\end{eqnarray}
These calculated intrinsic colors for different dust optical depths can then be compared to theoretically expected color indices of 
stars with different spectral type and luminosity class (e.g. Schmidt-Kaler
\cite{SK}), in order to constrain the range of possible spectral 
classifications.

The results are shown in Fig.\,\ref{lc_dust} for stars with 
luminosity class V, but the results for the other luminosity classes are very 
similar. Plotted are the dependences of the two intrinsic color indices on 
$\tau_{V}^{\rm dust}$ computed for the smallest and largest grains, as well as
for the MRN grain size distribution, using a mixture of silicates and 
amorphous carbon. The theoretically expected values
for the different spectral types are also shown. Obviously, an 
agreement between the computed and theoretical color indices (delivering 
the same dust optical depth) is achieved only for stars with spectral type 
between B6 and B7. Very similar results are found for the luminosity classes
III/IV, while for stars with luminosity class II we find a possible spectral
range of B8-8.5.

Interestingly, pure large grains are not able to account for the
circumstellar extinction. Instead, the circumstellar dust along the line of 
sight must consist of predominantly small grains, which can be located in an optically thin dust sphere or shell, in agreement with the results of Bjorkman et al. (\cite{Bjorkman}). However, other scenarios, like a dusty disk where the dust is not in our line of sight or is optically thin, since it is formed far from the star, cannot be discarded. An interferometric analysis will probably answer this question about the circumstellar dust geometry (Borges Fernandes et al., in preparation).

\subsubsection{The influence of the ionized circumstellar disk}\label{iondisk}

The contribution of an ionized envelope or disk to the total continuum emission
of early-type stars was the subject of many detailed investigations during the past 
years. For the case of classical Be stars, for instance, Zorec \& Briot 
(\cite{Zorec3}) found a relation between the excess emission in the $V$ band 
(defined as the difference between the $V$ band magnitudes measured during the 
Be phase and the non-Be phase) and the effective temperature of the star. Their 
relation indicates a stronger circumstellar contribution for hotter (i.e., 
earlier) Be stars, while late-type Be stars hardly show any contamination of their 
$V$ band fluxes by their circumstellar ionized material. A similar trend was 
found by detailed numerical studies of Stee \& Bittar (\cite{Stee}), who found 
circumstellar contributions to the continuum in the $B$ and $V$ 
bands for a B5e star of 0.57\% and 6.01\%, respectively, while the 
contributions are found to be much lower for later types. But not only are classical Be stars influenced by their ionized circumstellar material, B-type 
supergiants can also have a significant wind contribution at all optical bands, as recently shown by Kraus et al. (\cite{KKK}; \cite{KBK}).

B[e] stars are generally believed to have disks with higher densities than 
classical Be stars. In this case, an influence of the free-free and at optical 
wavelengths, especially of the free-bound continuum emission, might thus be 
expected. But in contrast to circumstellar dust, which acts at optical
wavelengths as a pure absorber that dims the stellar light, the free-free 
and free-bound processes act not only as an absorber, but at the same
time as an additional emission component, which adds more flux at a given 
wavelength than it absorbs from the underlying stellar flux. A star with an 
ionized disk thus usually appears brighter at all wavelengths compared
to a star with no ionized disk (see Kraus et al. \cite{KKK}; \cite{KBK}).

We check the influence of such an ionized circumstellar disk on the proper
spectral type determination by calculating the emission of free-free and 
free-bound processes in circumstellar disks of different densities.
The shape of the continuum emission of the ionized gas does not severely depend
on the geometry of the ionized material. For instance, a spherically symmetric wind and a geometrically flat disk result in the same shape of the wavelength dependent free-free and free-bound continuum 
emission. This indicates that the identical continuum flux distribution
can be obtained from quite different geometrical scenarios. In addition, the 
free-free and free-bound continuum at optical wavelengths is generated in the close 
vicinity of the star, typically within $2-5\,R_{*}$ (Kraus et al. \cite{KKK}). It can
thus not offer any insight into the global disk density distribution and geometry.

Since our aim is only to obtain proper stellar parameters, but not a global 
description of the detailed geometry and density variations within the highly 
dynamical and non-spherically symmetric circumstellar material, for which we 
definitely have too little observational information at hand, we study the influence
of the ionized disk emission on the color indices in a very simplified way. 

From the literature values (see Fig.\,\ref{ubvri}) we computed the pre-outburst
average observed magnitudes in the $UBV$ bands, deredden them with the 
interstellar extinction value derived in Sect.\,\ref{extinction}, and convert them 
into fluxes. These fluxes correspond to the star plus disk system. Then, individual
disk fluxes in the UBV bands were calculated. For this, we adopted a very
simple disk model whose surface density distribution can be described by
\begin{equation}
\Sigma(r) = \Sigma_{0} * \frac{R_*}{r}\, .
\end{equation}
Such a surface density distribution follows, e.g., for the case of an 
outflowing disk forming wind, which might be an appropriate model. In that 
case, the surface density at the inner edge of the disk is given by (see
Borges Fernandes et al. \cite{CD-42})
\begin{equation}
\Sigma_{0} = \frac{\dot{M} \alpha}{4 \pi R_* \varv} 
\end{equation}
where $\dot{M}$ describes the (here constant) mass loss rate over the 
disk-forming wind region, $\alpha$ is the disk opening angle, and 
$\varv$ the disk outflow velocity.

\begin{figure}[t!]
\begin{center}
\resizebox{\hsize}{!}{\includegraphics{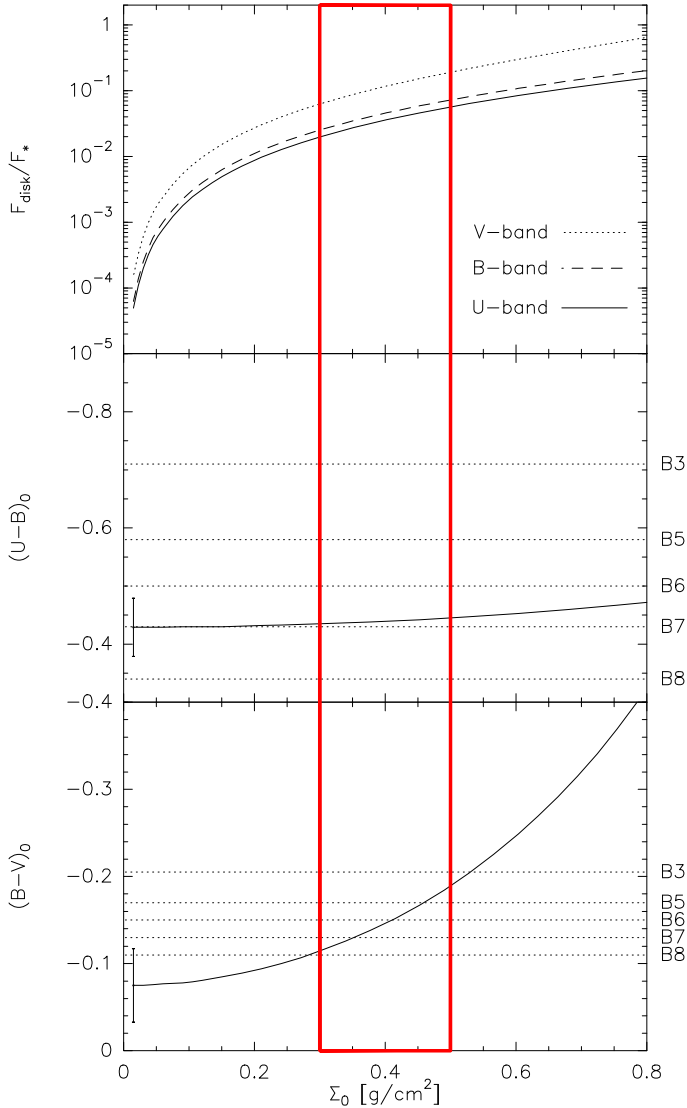}}
\caption{Increase in disk flux with respect to the stellar flux (top panel)
and intrinsic colors (mid and bottom panel) with disk surface 
density. The dotted 
lines indicate theoretically expected intrinsic colors for different spectral
types and a luminosity class V. The box extending over all three panels gives the range in possible 
classifications and the corresponding needed disk flux in all three bands.}
\label{disk}
\end{center}
\end{figure}

The disk fluxes (in the form of free-free and free-bound emission) in 
the $UBV$ bands, resulting from disks with increasing values of $\Sigma_{0}$, 
are then subtracted from the observed and interstellar extinction corrected ones. The
resulting pure stellar fluxes are converted back into magnitudes and the
color indices are derived. The results are shown in Fig.\,\ref{disk}.
Obviously, for low surface densities, the contribution of the disk is negligible,
while its importance drastically increases with increasing surface density
(top panel of Fig.\,\ref{disk}). 

Because of the increasing disk flux in all three bands with respect to the 
stellar flux, the star must be intrinsically fainter in all three bands (see Fig.\,\ref{disk}),
especially influencing the (B-V) color index (lower panel)
more than the (U-B) (middle panel). From a comparison of the derived color
indices to the expected intrinsic colors of Schmidt-Kaler (1982) for stars of luminosity class V and different spectral types, we find a rather narrow range of valid stellar
classifications, as indicated by the box extending over the complete
plot in Fig.\,\ref{disk}. While for luminosity classes III-V the range
in spectral type is B6-7, we find a spectral
type of B8 for luminosity class II. These ranges are in fairly good agreement with those found 
previously. This is not surprising, because the amount of circumstellar 
extinction is rather small, so that no big changes in intrinsic colors can be 
expected.
 
We would like to emphasize that the surface density chosen for the $x$-axis
to plot our results in Fig.\,\ref{disk} is not the most relevant parameter.
Instead, the influence of the disk as shown in the top panel is the crucial
parameter here. Such fractions of the free-free and free-bound continuum can 
also be achieved with quite different geometrical models and density
distributions. Nevertheless, the range in values for $\Sigma_{0}$ found from our analysis can easily be achieved with typical values for B[e] stars, e.g., disk opening angles in the range of 5 - 30 degrees, disk outflow velocities of 10 - 100 km\,s$^{-1}$, and mass loss rates typically of 10$^{-8}$ to 10$^{-5}$ $M_{\odot}$ yr$^{-1}$. However, the exact value for each parameter cannot be provided by our research due to the lack of reliable information concerning the geometry and dynamics of such a possible pre-outburst ionized disk, and the modeling provided here cannot be regarded as the only valid answer. It was only used to qualitatively discuss the influence of the ionized material to the color indices of the star in order to constrain the spectral classification of HD\,50138. How the ionized material is really distributed around the star cannot be derived from a pure free-free and 
free-bound emission calculation.

From our analysis of the circumstellar extinction contributions of either the dust
or the ionized disk, we find possible classifications of HD\,50138
as either a B6-7\,III-V star with $T_{\rm eff} = 13200\pm 500$\,K, or a 
B8-8.5\,II star with $T_{\rm eff} = 11300\pm 300$\,K. As a further check of these derived classifications, we derived the ratio of
the equivalent widths of the most plausible photospheric Si{\sc ii} lines
in our spectra. The ratios of the 4131\,\AA/4128\,\AA~and the 
5056\,\AA/5041\,\AA~equivalent widths are sensitive to temperature and surface gravity and have 
been computed from the line identification tables obtained with the code SYNSPEC 
(see Hubeny \& Lanz \cite{Hubeny}) based on Kurucz (\cite{Kurucz}) model atmospheres in local
thermodynamical equilibrium and Kurucz (\cite{Kurucz93}) line lists.
At temperatures between 10\,000\,K and 12\,000\,K and surface gravities according
to luminosity class II, these ratios show a mild increase with temperature
from 1.7 to 1.8 (4131\,\AA/4128\,\AA) and from 2.1 to 2.3 (5056\,\AA/5041\,\AA),
while the observed values, based on Table\,\ref{identification}, are $1.41\pm 0.19$ and $1.82\pm 
0.26$ in 1999 and $1.08\pm 0.15$ and $1.50\pm 0.21$ in 2007. From this discrepancy between observed and calculated equivalent width ratios, a B8-8.5 II scenario
for HD\,50138 can be excluded. For higher temperatures and especially higher 
surface gravities, the variations in the equivalent width ratios are no longer strictly correlated, but an overlap in the theoretical and observational values for stars
of spectral types in the range B6-7 III-V occurs, so that these equivalent width ratios can be
used at least as a qualitative check for consistency. From the photometric results
and the qualitative agreement with the equivalent width ratios of Si{\sc ii}
lines, we can claim that the final spectral classification of HD\,50138 is
a B6-7 III-V star. It is important to cite that our results agree with the classification obtained by Zorec et al. (\cite{Zorec}) and Cidale et al.(\cite{Cidale}) using the BCD (Barbier-Chalonge-Divan) spectrophotometric system, based on the study of the Balmer discontinuity.  

To finish our stellar parameter study, we want to determine the possible
ranges in stellar luminosities. Our ionized disk model implies a pure stellar $V$ band flux from 6.49\,mag to 6.71\,mag for the lowest and highest possible disk contributions of 7\% and 20\%, respectively, in the range of possible MK types. We thus conclude that the star has an intrinsic $V$ band flux 
of $6.60 \pm 0.11$\,mag. This result has been obtained from considering the 
pre-outburst photometry only. We also tried it with the post-outburst data. 
Thanks to the high variability of the data, the results still agree with 
a B6-7 star but have a higher uncertainty. Using the bolometric corrections 
for B6-7\,III-V stars from Flower (1996) and the Hipparcos distances 
towards HD\,50138 of $500\pm 150$\,pc, we can finally 
calculate the stellar luminosity to $\log(L_*/L_{\odot}) = 3.06 \pm 0.27$.

We did not attempt to derive the luminosities for the case of the circumstellar 
dust extinction, because (i) based on the observed ionized disk around HD\,50138, 
this possibility seems to be the more realistic one, and (ii) we currently have no clear indication of the possible dust composition and grain size distribution 
at hand. Nevertheless, assuming the range in dust optical dephts obtained by our 
analysis for the MRN grain size distribution (see Fig.\,\ref{lc_dust}), the range of stellar luminosities corrected by the circumstellar dust extinction is in a fairly good agreement with the above derived luminosity range.

\section{Discussion of the nature of HD\,50138}\label{discussion}

The variability in the line profiles, as cited by Pogodin (\cite{Pogodin}), seems to be caused by shell phases and/or outburst events. As described in the Sect.\,\ref{spectral_descr}, from comparison of our spectroscopic data, we saw that there is a sort of temporal evolution, in particular an increase of a blue emission component and a redshift of the absorption ones. Based on this, we believe that a new ejection of material took place prior to our Narval observations, i.e., before March 2007 with increasing emission later on (our second FEROS observation). This new shell phase would be the responsible for the changes seen in our data. 

Another possibility is that the material was ejected in our direction, possibly by a hot spot on the stellar surface, forming a kind of ``one-armed spiral", as suggested for classical Be stars. If this was the case, this would mainly explain the blue-shifted emission. On the other hand, even without having complete knowledge of the wind contribution for all lines, we suggested that for some of them, there is no contamination from the wind and they have a pure photospheric origin. Based on this, an extended and expanding atmosphere or a new ring or envelope with a lower outflow velocity that started to expand some time in early 2007 (or slightly earlier) could explain the changes in the radial velocities seen in the photospheric lines {(see Fig.\,\ref{Si_ii} and Table\,\ref{identification}).}

The results of Bjorkman et al. (\cite{Bjorkman}), claiming the existence of intrinsic polarization due to electron scattering, indicate a non-spherically symmetric structure with a gaseous disk seen almost edge-on, in association with an optically thin dust envelope. Our results described in Sect.\,\ref{dust} confirm that such a scenario for the circumstellar dust is possible. 

\begin{figure}[t!]
\begin{center}
\resizebox{\hsize}{!}{\includegraphics{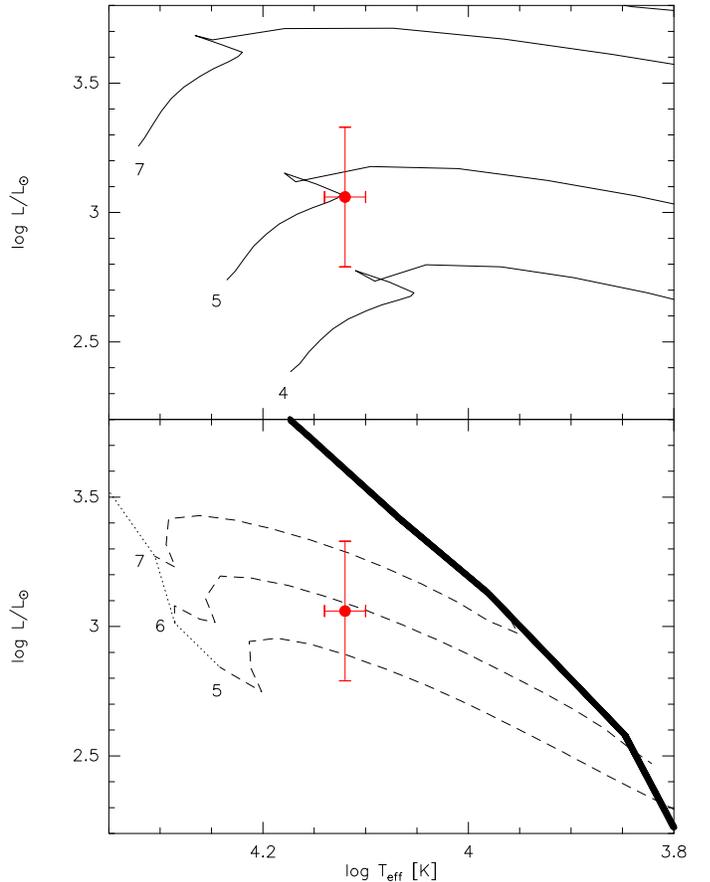}}
\caption{Position of HD\,50138 on the HR diagram compared to 
post-main sequence (top panel) and pre-main sequence (bottom panel)
evolutionary tracks. In the bottom panel, the dotted line defines the 
zero-age main sequence, the thick solid line defines the birthline. }
\label{hrd}
\end{center}
\end{figure}

Based on the stellar parameters derived in Sect.\,\ref{iondisk}, we can indicate
the position of HD\,50138 on the HR diagram and compare it to evolutionary tracks. 
This is shown in Fig.\,\ref{hrd}, where we plot the evolutionary 
tracks from Schaller et al. (\cite{Schaller}) in the top panel for stars at solar metallicity. From 
this plot, HD\,50138 is a $(5.0\pm 0.5)\,M_{\odot}$ star, since still on, or just 
evolving off, the main sequence. The lower panel of Fig.\,\ref{hrd} shows its position compared to pre-main sequence 
evolutionary tracks from Bernasconi \& Maeder (\cite{Bernasconi}). The
ZAMS, as well as the birthline are also included.
This plot indicates a stellar mass of roughly $(6.0\pm 1.0)\,M_{\odot}$, 
but for a star that is approaching the main sequence.

From the position in the HR diagram alone, it is not possible to distinguish
between a pre-main sequence and a post-main sequence evolutionary phase, however we can discuss some points about the possible nature of HD\,50138. 

\begin{itemize}

\item Pre-main sequence nature:

\end{itemize}

The luminosity and the mass derived by our analysis agree with those expected for an intermediate mass pre-main sequence star. HD\,50138 shows similar spectral and photometric variations as seen in young objects. However, if it is a young star close to the main sequence, associated to its high temperature, its maternal cloud would already be dispersed and the remaining dust would be located in the accretion disk. This would be confirmed by the forbidden lines seen in our spectra, which are not blueshifted, as seen in young deeply embedded stars (Waters \& Waelkens \cite{Waters}; Hamann \cite{Hamann}). In addition, based on its rather high temperature, HD\,50138 would be a Herbig B[e] star.

On the other hand, the absence of any nebulosity around HD\,50138 and its situation far from any star-forming region (Pogodin \cite{Pogodin}) is against a Herbig B[e] classification. The possible shell ejection phases are also against a young nature. In addition, the results of Bjorkman et al. (\cite{Bjorkman}) show that the dust is probably distributed in an optically thin envelope and not in an accretion disk. 

\begin{itemize}

\item A Be star close to (or just at the turn-off from) the main sequence:

\end{itemize}

The classification as a Be star that is still on, or just at the turn-off from, the main sequence cannot be ruled out at this moment. However, two points are against the classification of HD\,50138 as a classical Be star: the forbidden lines and the IR excess due to dust. The formation of dust in an intermediate-mass star close to the main sequence seems to be a key problem for this scenario. 

The high-density circumstellar matter and the gaseous ionized disk-like structure 
speak for rather high mass loss. This mass loss must have been much higher than in the case of classical Be stars, forming disk-like or spiral arm structures, if the material was expelled from some hot spots on the stellar surface and expanding outwards. Because of the much higher initially released masses in equatorial direction, the density in the outer regions will have ideal conditions for the production of the forbidden emission lines. This mass loss might be less than in the case of B[e] stars. Then, in agreement with Jaschek et al. (\cite{Jaschek2}), we might claim that HD\,50138 is {\it ``a transition object between Be and B[e] stars"}.

\begin{itemize}

\item Binary system:

\end{itemize}

For such a star to have a large amount of dust, as seen in its SED, another possible scenario would be linked to a binary component, where the dust could be formed in a long-lived circumbinary disk or by wind collisions. Baines et al. (\cite{Baines}) suggest that the (possible) separation of components would be at least 0.\arcsec5. However, considering the possible distance of HD\,50138, this separation would be very large to have mass exchange (250\,AU considering the smallest angular separation), which could explain the spectral and photometric variability of this object. This would also hamper the possible classification of this star as FS CMa star (Miroshnichenko \cite{Miroshnichenko}), since this group of stars is tentatively classified as objects close to or still on the main sequence in binary systems with mass exchange.

On the other hand, since the radius of the FEROS aperture is 1\arcsec, in principle, there might be a chance that we observed both components at the same time. However, we could not find any spectral feature to confirm the existence of a secondary component. This result also agrees with Corporon \& Lagrange (\cite{Corporon}), who searched for TTauri companions in a sample of HAeBe candidates, based on several diagnostic lines, e.g. from Ca{\sc i}, Fe{\sc i} and Li{\sc i}. They did not find any evidence of binarity for HD\,50138. However, an HAeBe star as a companion could not be completely discarded. Thus, at this moment we can neither confirm nor exclude a possible binary nature for HD\,50138. 

\begin{itemize}

\item Other classes of B[e] stars:

\end{itemize}

Since HD\,50138 shows the B[e] phenomenon, we should also discuss the possibility that it belongs to one of the other known classes of objects that present it (Lamers et al. \cite{Lamers}). However, based on the absence of typical nebular and symbiotic lines in its spectrum, we can discard a compact planetary nebula and a symbiotic nature for this object. In addition, the low luminosity, in combination with no indication of nitrogen enrichment, exclude a supergiant scenario for this star. However, since almost 50\% of the objects with the B[e] phenomenon are considered unclassified, we cannot exclude that HD\,50138 might be an object that represents a link between Be and B[e] stars.

\section{Conclusions}\label{conclusion} 

HD\,50138 is a very curious star that displays strong spectroscopic variations. Based on analysis of new high-resolution data, we present a detailed description of these variations. Our analysis of the photometric data suggests that HD\,50138 is a B6-7 III-V star, whose luminosity was tentatively obtained from a careful study of the influence of the possible circumstellar extinction sources and based on the new Hipparcos 
distance. A new value for the color excess, $E(B-V) = 0.08\pm 0.01$\,mag, was derived.
In addition, we suggest that a new-shell phase or the formation of one-armed spiral could have taken place before 2007. 

Based on our results, a pre-main sequence star or a transition object between Be and B[e] stars, close to or just at the turn-off from the main sequence, or a binary scenario, can be neither confirmed nor discarded; however, an observational campaign, based on photometry, to derive a detailed light-curve and high-resolution spectroscopy associated to a detailed analysis in terms of the line profile appearances and variations have to be performed to confirm the possible shell phases and the slow-down of this star. A careful interferometric analysis, associated to a SED modeling considering different scenarios for the circumstellar dust is in progress (Borges Fernandes et al., in preparation) and will certainly provide better constraints for the circumstellar geometry and the nature of this curious star.


\begin{acknowledgements}
This research made use of the NASA Astrophysics Data System (ADS) and of the
SIMBAD, VizieR and 2MASS databases. M.B.F. acknowledges financial support from the Programme National de Physique Stellaire (France) and the Centre National de la Recherche Scientifique (CNRS-France) for the post-doctoral grant. M.K. acknowledges financial support from GA\,AV \v{C}R number KJB300030701. M.B.F. and F.X.A. acknowledge Dr. Victor de Amorim d'\'Avila to perform the observations at ESO (La Silla, Chile). A.D.S. acknowledges Dr. Michel Auriere to perform the observations at the Observatoire Midi Pyrenees (France). M.B.F. acknowledges Dr. Pierre Cruzal\'ebes for his help with Hakkila et al. (1997) software. We acknowledge Dr. Ad\'ela Kawka for comments on the manuscript and we also wish to thank Dr.\,Ji\v{r}\'i Kub\'at for providing us with his databases computed with SYNSPEC.
\end{acknowledgements}


\end{document}